\documentclass{emulateapj}
\usepackage{graphicx}
\usepackage{hyperref}
\usepackage{color}

\newcommand{\ergs}{${\rm erg \ cm^{-2} \ s^{-1}}$ }

\newcommand{\todo}{\ifmmode {\Huge \bullet} \else {\Huge$\bullet$}\fi}

\newcommand{\vFWHM}{\ifmmode V_{\mbox{\tiny FWHM}} \else $V_{\mbox{\tiny FWHM}}$ \fi}

\newcommand{\kms}{\ifmmode {\rm km\,s}^{-1} \else km\,s$^{-1}$ \fi}
\newcommand{\cc}{\hbox{cm$^{-3}$}}
\newcommand{\cmii}{\hbox{cm$^{-2}$}}
\newcommand{\ergcms}{\ifmmode {\rm ergs\,cm}^{-2}\,{\rm s}^{-1} \else ergs\,cm$^{-2}$\,s$^{-1}$\fi}
\newcommand{\ergcmsA}{\ifmmode{\rm ergs}\, {\rm cm}^{-2}\,{\rm s}^{-1}\,{\rm\AA}^{-1} \else ergs\, cm$^{-2}$\, s$^{-1}$\, \AA$^{-1}$\fi}
\newcommand{\ergcmsHz}{\ifmmode{\rm ergs\,cm}^{-2}\,{\rm s}^{-1}\,{\rm Hz}^{-1} \else ergs\,cm$^{-2}$\,s$^{-1}$\,Hz$^{-1}$\fi}
\newcommand{\phcms}{\ifmmode {\rm ph\,cm}^{-2}\,{\rm s}^{-1} \else ,ph\,cm$^{-2}$\,s$^{-1}$\fi}
\newcommand{\phcmsA}{\ifmmode {\rm ph\,cm}^{-2}\,{\rm s}^{-1}\,{\rm\AA}^{-1} \else ph\,cm$^{-2}$\,s$^{-1}$\,\AA$^{-1}$\fi}

\newcommand\Msun{\ifmmode M_{\odot} \else $M_{\odot}$\fi}
\newcommand\msun{\ifmmode M_{\odot} \else $M_{\odot}$\fi}
\newcommand\Lsun{\ifmmode L_{\odot} \else $L_{\odot}$\fi}

\newcommand\mpyr{\ifmmode \Msun\,{\rm yr}^{-1} \else $\Msun\,{\rm yr}^{-1}$ \fi}

\newcommand{\Luv}{\ifmmode L_{1450} \else $L_{1450}$\fi}
\newcommand{\Lop}{\ifmmode L_{5100} \else $L_{5100}$\fi}
\newcommand{\Lthree}{\ifmmode L_{3000} \else $L_{3000}$\fi}
\newcommand{\lledd}{\ifmmode L/L_{\rm Edd} \else $L/L_{\rm Edd}$\fi}
\newcommand{\ledd}{\ifmmode L_{\rm Edd} \else $L_{\rm Edd}$\fi}
\newcommand{\lamLlam}{\ifmmode \lambda L_{\lambda} \else $\lambda L_{\lambda}$\fi}
\newcommand{\lbol} {\ifmmode L_{\rm Bol} \else $L_{\rm Bol}$\fi}
\newcommand{\llbol}{\ifmmode \log\left(\lbol/\ergs\right) \else $\log\left(\lbol/\ergs\right)$\fi}

\newcommand{\fuv}{\ifmmode f_{\lambda}\left(1450\AA\right) \else $f_{\lambda}\left(1450 {\rm \AA}\right)$\fi}
\newcommand{\fthree}{\ifmmode f_{\lambda}\left(3000\AA\right) \else $f_{\lambda}\left(3000{\rm \AA}\right)$\fi}
\newcommand{\fH}{\ifmmode f_{\lambda}\left(1.65\micron\right) \else
$f_{\lambda}\left(1.65\micron\right)$\fi}

\newcommand{\mbh}{\ifmmode M_{\rm BH} \else $M_{\rm BH}$\fi}
\newcommand{\lmbh}{\ifmmode \log\left(\mbh/\Msun\right) \else $\log\left(\mbh/\Msun\right)$\fi}

\newcommand \Hbeta {\ifmmode {\rm H}\beta \else H$\beta$\fi}
\newcommand \hb    {\ifmmode {\rm H}\beta \else H$\beta$\fi}
\newcommand  \mgii  {\ifmmode {\rm Mg}{\textsc{ii}} \else Mg\,{\sc ii}\fi}
\newcommand  \MGII  {\ifmmode {\rm Mg}\,{\sc ii}\,\lambda2798 \else Mg\,{\sc ii}\,$\lambda2798$\fi}
\newcommand  \siiv  {\ifmmode {\rm Si}\, {\sc iv}\ \else Si\,{\sc iv}\fi}
\newcommand  \SIIV  {\ifmmode {\rm Si}\,{\sc iv}\,\lambda1399 \else Si\,{\sc iv}\,$\lambda1399$\fi}
\newcommand  \civ  {\ifmmode {\rm C}\, {\sc iv}\ \else C\,{\sc iv}\fi}
\newcommand  \CIV  {\ifmmode {\rm C}\,{\sc iv}\,\lambda1549 \else C\,{\sc iv}\,$\lambda1549$\fi}
\newcommand  \NV  {\ifmmode {\rm N}\,{\sc v}\,\lambda1240 \else N\,{\sc v}\,$\lambda1240$\fi}
\newcommand  \nv  {\ifmmode {\rm N}\,{\sc v}\ \else N\,{\sc v}\fi}
\newcommand  \LyA  {\ifmmode {\rm Lyman}\,{\sc $\alpha$}\,\lambda1216 \else Lyman\,{\sc $\alpha$}\,$\lambda1216$\fi}
\newcommand  \lya {\ifmmode {\rm Lyman}\,{\sc $\alpha$}\ \else Lyman\,{\sc $\alpha$}\fi}
\newcommand  \feii     {\ifmmode {\rm Fe}\,{\sc II}\,\lambda1785.4 \else Fe\,{\sc II}\,$\lambda1785.4$\fi}

\newcommand  \aliii  {\ifmmode {\rm Al}{\textsc{iii}} \else Al\,{\sc iii}\fi}
\newcommand  \ALIII  {\ifmmode {\rm Al}\,{\sc iii}\,\lambda1857 \else Al\,{\sc iii}\,$\lambda1857$\fi}

\newcommand  \CIII  {\ifmmode {\rm C}\,{\sc iii]}\,\lambda1909 \else C\,{\sc iii]}\,$\lambda1909$\fi}
\newcommand  \oi    {\ifmmode \left[{\rm O}\,{\textsc i}\right] \else [O\,{\sc i}]\fi}
\newcommand  \OI    {\ifmmode \left[{\rm O}\,{\textsc i}\right]\,\lambda6300 \else [O\,{\sc i}]$\,\lambda6300$ \fi}

\newcommand  \oii   {\ifmmode \left[{\rm O}\,{\textsc ii}\right] \else [O\,{\sc ii}]\fi}
\newcommand  \OII   {\ifmmode \left[{\rm O}\,{\textsc ii}\right]\,\lambda3727 \else [O\,{\sc ii}]\,$\lambda3727$ \fi}
\newcommand  \oiii  {\ifmmode \left[{\rm O}\,{\textsc iii}\right] \else [O\,{\sc iii}]\fi}
\newcommand  \OIII  {\ifmmode \left[{\rm O}\,{\textsc iii}\right]\,\lambda5007 \else [O\,{\sc iii}]\,$\lambda5007$\fi}

\newcommand{\lmg}{\ifmmode L\left(\mgii\right) \else $L\left(\mgii\right)$\fi}
\newcommand{\fwmg}{\ifmmode {\rm FWHM}\left(\mgii\right) \else FWHM(\mgii)\fi}
\newcommand{\fwciv}{\ifmmode {\rm FWHM}\left(\civ\right) \else FWHM(\civ)\fi}
\newcommand{\fwhm}{\ifmmode {\rm FWHM} \else FWHM\fi}

\begin{document}
%\title{Statistical evidence for Photoionization Dominant Broad Absorption Line Variability}
%\title{Photoionization dominates the Broad Absorption Line Variability}
\title{Variation of ionizing continuum: the main driver of Broad Absorption Line Variability}
%What Drives Quasar Absorption Line Variability?}
\author{Zhicheng He\altaffilmark{1}, Tinggui Wang\altaffilmark{1}, Hongyan Zhou\altaffilmark{1}, Weihao Bian\altaffilmark{2}, Guilin Liu\altaffilmark{1}, Chenwei Yang\altaffilmark{1}, Liming Dou\altaffilmark{1}, and Luming Sun\altaffilmark{1}} 
\altaffiltext{1}{
CAS Key Laboratory for Research in Galaxies and Cosmology, Department of 
Astronomy, University of Science and Technology of China, Hefei, Anhui 
230026, China; zcho@mail.ustc.edu.cn}
\altaffiltext{2}{
Department of Physics and Institute of Theoretical Physics, Nanjing Normal University, Nanjing 210023, China}
\begin{abstract}
We present a statistical analysis of the variability of broad absorption lines (BALs) in quasars using the large multi-epoch spectroscopic dataset of the Sloan Digital Sky Survey Data Release 12 (SDSS DR12). We divide the sample into two groups according to the pattern of the variation of \civ~BAL with respect to that of continuum: the equivalent widths (EW) of the BAL decreases (increases) when the continuum brightens (dims) as group T1; and the variation of EW and continuum in the opposite relation as group T2. We find that T2 has significantly ($P_{\rm T}<10^{-6}$, Student’s T Test) higher EW ratios (R) of \siiv~to \civ~BAL than T1. Our result agrees with the prediction of photoionization models that $C^{+3}$ column density increases (decreases) if there is a (or no) $C^{+3}$ ionization front while R decreases with the incident continuum. We show that BAL variabilities in at least 80\% quasars are driven by the variation of ionizing continuum while other models that predict uncorrelated BAL and continuum variability contribute less than 20\%. Considering large uncertainty in the continuum flux calibration, the latter fraction may be much smaller. When the sample is binned into different time interval between the two observations, we find significant difference in the distribution of R between T1 and T2 in all time-bins down to a $\Delta T<6$ days, suggesting that BAL outflow in a fraction of quasars has a recombination time scale of only a few days.
\end{abstract}


\section{Introduction}

\label{sec:intro}
Quasar outflows have been considered as a primary feedback source
to explain the co-evolution of the central supermassive black holes
(SMBHs) and their host galaxies. Theoretical studies and simulations
suggested that kinetic power of order of only ~1\% of the Eddington
luminosity is deemed sufficient for significant feedback effects on
the host galaxy (e.g., Scannapieco \& Oh 2004; Di Matteo , Springel
\& Hernquist 2005; Hopkins et al. 2006; Moll et al. 2007;
Hopkins \& Elvis 2010). Furthermore, in the past ten years, UV and
X-ray spectroscopic studies have revealed powerful high-velocity
outflows carrying kinetic power of at least a few percent of bolometric
luminosity in some bright active galactic nuclei (AGNs) and quasars
(e.g., Borguet et al. 2012; Tombesi 2010, 2011). Finally, massive molecular
outflows on the galactic scale were detected in infrared luminous quasars,
which is thought in the early phase of nuclear activity, with velocities
correlated with the AGN luminosity (e.g., Sturm 2011).

Outflows may manifest themselves as blue-shifted absorption lines
when they intercept the line-of-sight to the continuum source. Broad
absorption lines (BALs) up to velocity in 0.2 c from the correspondent
emission lines are seen in about 10-40\% of quasar population (e.g. Gibson
et al. 2009; Allen et al. 2011). Technically, they are defined as absorption
troughs with velocity widths $>$2000 \kms~at depths $>$10 per cent below
the continuum (Weymann et al. 1991). Most BAL quasars display absorption
lines of only high-ionization species, such as \NV, \CIV~and \SIIV, known
as HiBAL (Weymann et al. 1991; Gibson et al. 2008; 
Filiz Ak et al. 2013), 
and a few percent also of low ionization species such as \MGII~
and \ALIII, named as LoBAL (Weymann et al. 1991; Zhang et al.
2010). However, mass outflow rate and the kinetic
power associated with BAL outflows are poorly determined due to lack of
density diagnostics or the distance to the central black hole.

BAL troughs are variable on time scales from several days to
years, and the fraction of variable BALs increases with increasing
observing intervals (Gibson et al. 2008; Capellupo et al. 2011; Wildy et al. 2014).
So far, two mechanisms are invoked to explain the variability of BAL
troughs: (1) changes in the ionization of gas; and (2) absorbing gas
moving in and out of the line-of-sight. In the first case, the alteration
of gas ionization can be caused by a varying incident ionizing
continuum (case 1a) or the evolving gas density (case 1b) due
to, e.g., thermal instability (Waters et al. 2016). 
In either case, the variability of BAL troughs can provide either
important clues to origin or crucial diagnostics for the physical
conditions of outflows (Lundgren 
et al. 2007; Gibson et al. 2008, 2010; Filiz Ak et al. 2012,
2013; Capellupo et al. 2012, 2013; Welling et al. 2014;
He et al. 2014, 2015; Wang et al. 2015). 
For example, in case 1a), the variability time scale will set an
upper limit on recombination time scale, which in turn gives a lower
limit on the gas density. In the second case, assuming a Keplerian
dominated transverse motion for the absorber, we can constrain the
distance of the absorber to the central black holes using the variability 
time scale of the BAL troughs (Moe et al. 2009; 
Capellupo et al. 2011). In order to use these constraints,
a first step is to establish which mechanism is responsible
for the BAL variability.

These mechanisms predict very different relations between continuum
and absortpion line variabilities. In case 1a), variability of continuum
and absorption lines are highly correlated; while in other two cases,
they are independent (see Waters et al. 2016 for thermal instability
case). Furthermore, along the line sight, different part of outflows
see the same ionizing continuum, thus different absorption components
vary also coordinately in the case 1a), while in other two cases, we do
not expect such coherent variations. Previous studies showed that the
same BAL troughs of \civ~and \siiv~BALs or different BAL components of
\civ~in one object vary in a coordinated manner (strengthening or
weakening) (e.g. Filiz Ak et al. 2012, 2013; Wildy et al. 2014),
indicating that ionization change due to varying ionizing continuum
plays important role (Hamann et al. 2011; Wang et al. 2015). It
is inconclusive whether variations of BAL EW correlates with continuum
flux in individual quasars (as found in e.g, Trevese et al. 2013; 
Capellupo, Hamann \& Barlow 2014).
Wang et al. (2015, hereafter paper I) presented a statistical study
of ensemble correlation between BALs and continuum for a large sample
of variable BAL quasars with multi-epoch spectroscopic dataset of the
SDSS DR10. They found that the EWs of the BALs decrease (increase)
statistically when the continuum brightens (dims) for about 70\%
BALs, and concluded that varying ionizing continuum drove BAL
variability in most of these quasars. They also showed that emergence
or disappearance of BALs are also connected to the continuum variability.  

However, a number of issues have not been addressed in that paper.
First, the coordinated variations of emission and continuum are also
expected in the scenario of cloud moving in and out picture if BAL
outflows carry dust as suggested by some studies (Zhang et al. 2015;
He et al. 2014; Leighly et al. 2015). Second, the relation between
BAL and continuum variability is not monotonic according to
photoionization models (see their Figure 11 and also Figure 3 in
\S3). Consequently, it lowers the ensemble correlation between BAL
and continuum variability. A quantitative analysis should take into
account of this. 

In this paper, we will examine the role of variation of
continuum playing in the BAL variabilities and give a
quantitative estimation for the fraction of BAL variability 
driven by the photoionization. The paper is organized
as follows. We describe the sample and data analysis in
\S2. An indicator of the ionization state is described in \S3.
The results are presented in \S4 and \S5. The conclusions
are given in \S6.

\section{Variable Absorption Line Quasar Sample}
\label{sec:sample}
We merged the BAL quasar in the catalog of SDSS data release 7 (DR7, Shen et al. 2011)
\footnote{http://das.sdss.org/va/qso\_properties\_dr7/dr7.htm} with that of DR12 by P\^{a}ris et al. 2016
\footnote{http://dr12.sdss3.org/datamodel/files/BOSS\_QSO/DR12Q/
DR12Q\_BAL.html}. 
Duplicated entries are removed.  We compared this catalog with the SDSS
spectroscopic catalog and selected quasars with multi-spectroscopic observations. To investigate
the variability of \civ~and \siiv, we adopt a redshift cut 1.9 $<$ z $<$ 4.7. To ensure detection
of major absorption lines, we only keep quasars with at least one spectrum with signal to noise
SNR $>$10. After these cuts, we obtain a sample of 2005 BAL quasars with two spectra or more.
For one quasar with m $(m\geq2)$ spectra, there are $C^2_{m} = m(m-1)/2$ spectra pairs.
As a result, there are total 9918 spectrum pairs in the sample of 2005 BAL quasars.
In the following subsection we describe the method of construction of a variable absorption line
quasar sample.
\subsection{Spectra flux calibration}
One of the primary scientific goals of Baryon Oscillation Spectroscopic Survey (BOSS) is 
to search for the baryon acoustic oscillation (BAO) signal in the \lya~forest. As mentioned in Paper I, 
the fiber positions of BOSS quasar targets were purposefully offset in order to optimize the throughput 
of light at 4000\AA, while the standard stars used for flux calibration are positioned for 5400\AA~
(Dawson et al. 2013). As a result, there exists the systematic offest between the spectrophotometric 
flux calibrations of SDSS and BOSS. 

The quasar locus for high redshift quasars overlaps with the stellar locus in SDSS color-color space. 
As a result, many stars are targeted as quasars until the spectra were taken. These 
stars were observed in the same manner as the observed quasars (Harris et al. 2016). 
Thus, they have the same systematic offset with the quasars in BOSS. 
We give an average correction to the BOSS quasar spectra. The correction is small at the bluest wavelength
and up to about 17\% at the reddest wavelength (Fig. 13 of Harris et al. 2016).

%\begin{figure}
%\center{}
%\includegraphics[height=6.4cm]{bosstosdss.eps}
%\caption{The ratio between BOSS "stellar contaminants" spectra and their counterparts
%in SDSS (Harris et al. 2016).}
%\end{figure}

\subsection{Using unabsorbed Quasar Templates to fit the spectra}
In order to determine the wavelength span of the \civ~and \siiv~BAL trough, we fit the spectra using
unabsorbed template quasar spectra as in Paper I, which are
drived from 38,377 non-BAL quasars with $1.5 < z < 4.0$ and $SNR_{\rm 1350} > 10$ in SDSS Data Release 7 (DR7).
Like Paper I, we fit these templates to the spectra using a double power-law function as a scale factor,
\begin{equation}
S_\lambda=A[1]\left(\frac{\lambda}{2000\AA}\right)^{A[2]}+ 
    A[3]\left(\frac{\lambda}{2000\AA}\right)^{A[4]}
\label{eq1}
\end{equation}
where coefficients (A[1],A[3]) and exponents(A[2],A[4]) are determined by minimizing $\chi^2$. This fitting spans a wavelength range of
1150 to 2850\AA\ and we add a additional Gaussian component at the \lya, \nv, \siiv~and \civ~locations to the unabsorbed templates to improve the fits therein. We iteratively mask spectral pixels lower than the model with a significance of $\ge 3\sigma$ on the blue side of \lya, \nv, \siiv, \civ~to exclude possible absorption lines. 
We select the best template according to the criterion of the maximum number of pixels within 1$\sigma$ error
as the final fitting. The fitting procedure is demonstrated in Fig. 1.

Like Paper I, we search contiguous deficient pixels for intrinsic absorption lines region of \civ~in 
the normalized spectrum. We screen the moderate to broad absorption lines, 
which are more likely intrinsic, deficiency over a width of $\Delta \ln\lambda\geqslant 
10^{-3}$, or 300 km~s$^{-1}$ in velocity, and statistically significant 
than 5$\sigma$. Then we check by eye to exclude the false ones, caused by an improper fit in most cases. 
The absorption line regions of \siiv~are defined according to the corresponding velocity of \civ.

\begin{figure*}
\center{}
\includegraphics[height=9.5cm]{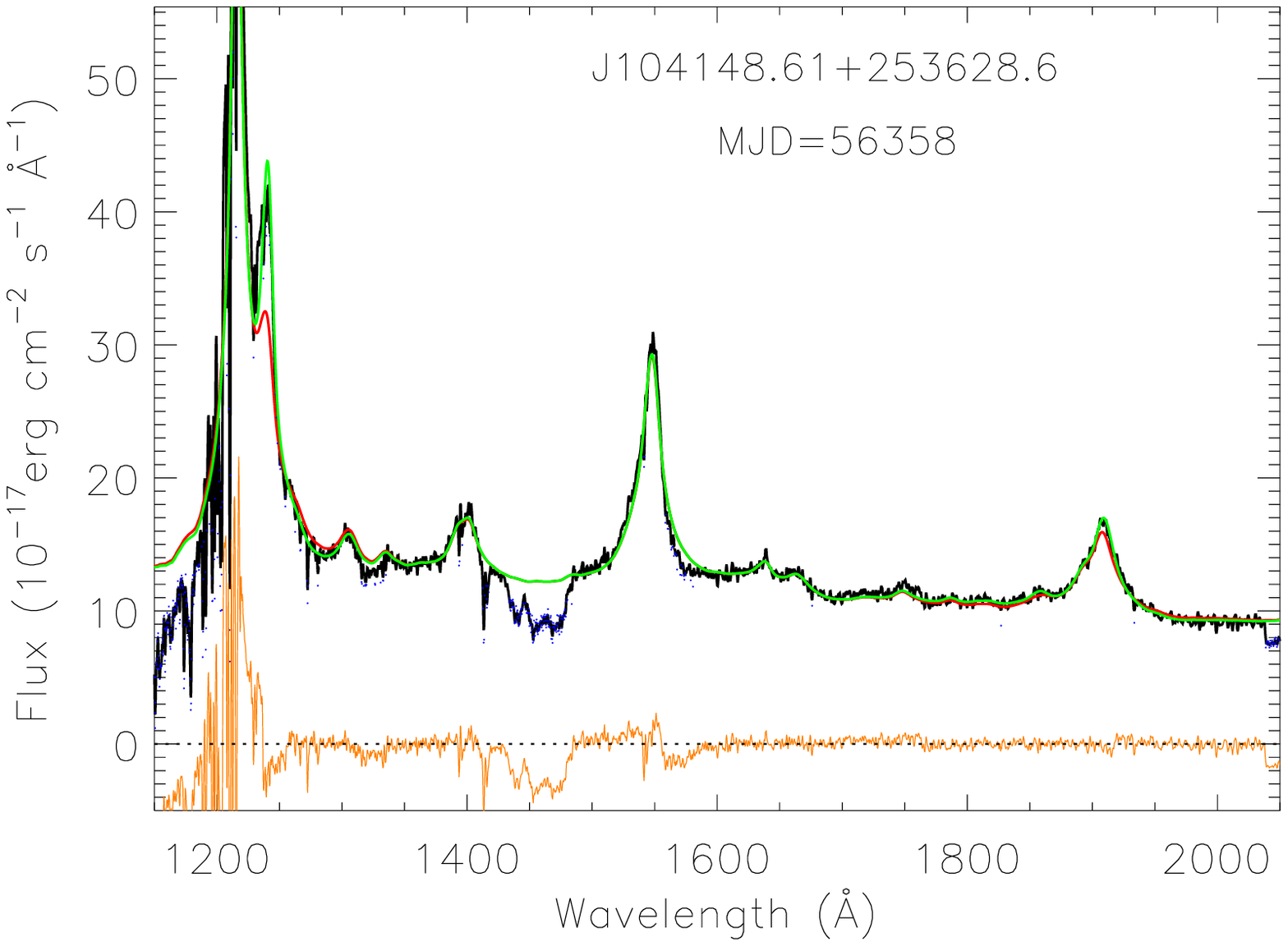}
\includegraphics[height=9.5cm]{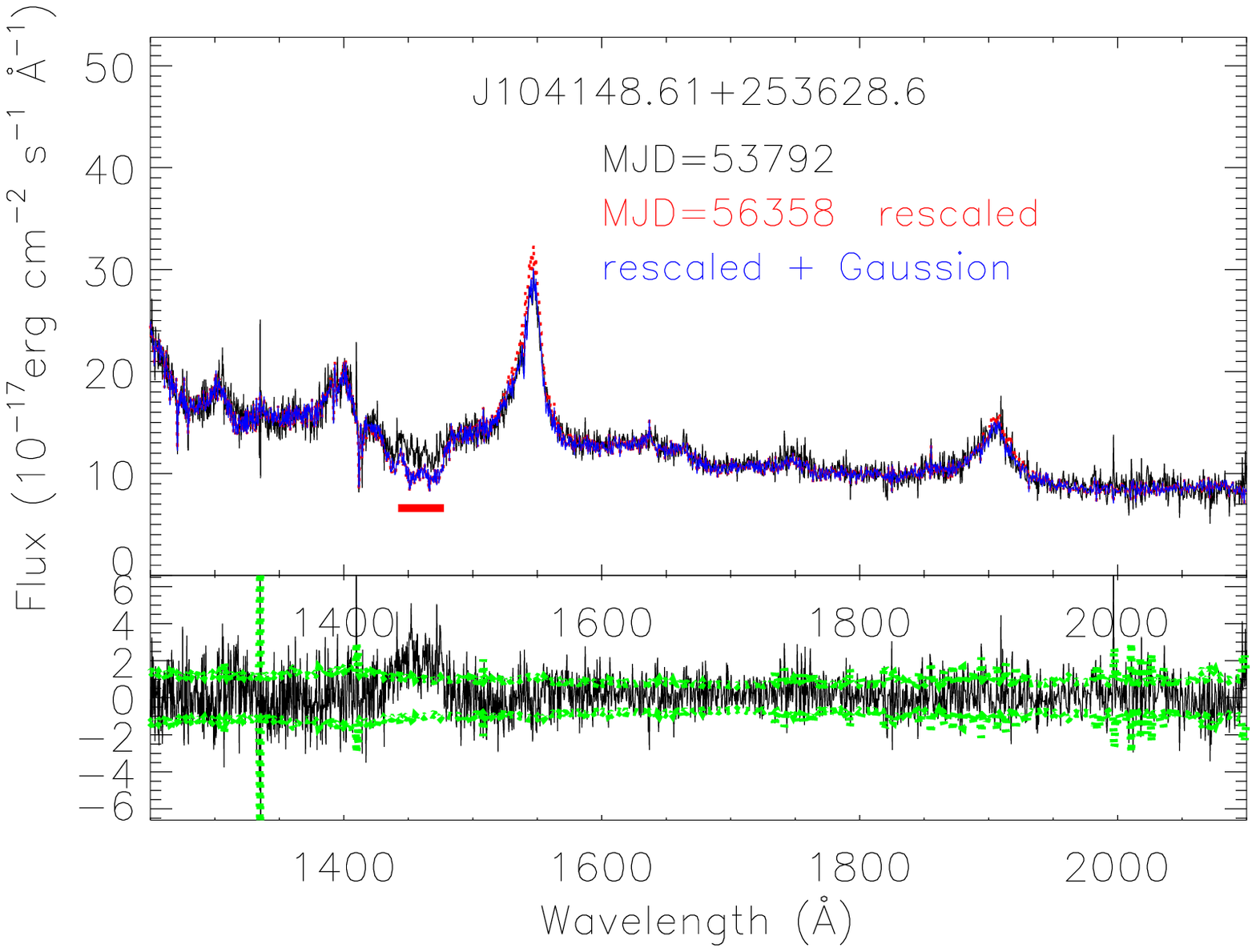}
\caption{Top panel: an example of unabsorbed template fit to the spectra of the SDSS quasars. The red 
line shows the scaled best matched template, while the green line has additional gaussians to 
account for emission lines. The blue points are those pixels $3\sigma$ below the fit, 
and thus masked out during the fitting. The orange line is residual spectra. 
Bottom panel: Example of matching the reference spectrum to another spectrum (in black) by multiplying the 
reference spectrum with a double power-law described in the text (the red line). The blue curve represents 
the one with additional Gaussians to account for the 
change of the emission line equivalent width. The red horizontal line represents the varied region of \civ~BAL.
The residuals of fits (solid line) and the combined spectrum 
uncertainties (dashed line) are plotted in the lower panel.}
\end{figure*}

\subsection{Identification of Variable Absorption Lines}
As described in Paper I, in order to identify the variation region of \civ~BAL, we select the higher $SNR$ 
spectrum of the pair of a quasar as a reference, and then rescale it using the 
double power-law function (Eq.\ref {eq1}) to account for potential variations of the continuum shape.  
As shown below, with this recipe we can fit the observations very well. Examples of the 
fit are displayed in the Figure 1. 
To account for variations of emission line EW, we add/subtract 
a Gaussian to/from the rescaled spectrum. We restrict the center of the Gaussian to lie 
within $\pm$500 km/s of the line center at the source rest frame.
The add/subtract Gaussian will be used as an indicator of the 
increase or decrease of \civ~emission line EW. With this fit, we found that the 
scatter in the difference spectrum over emission-line regions is similar to that 
of continuum regions. Compared with unabsorbed quasar template 
matching, the rescaled reference matching usually produces a better fit outside 
the absorption line region. As a result, we will measure the absorption 
line variability from the difference spectrum rather than by comparing EWs 
obtained in the template fitting.

To identify variable absorption-line components in the difference spectrum, we 
searched for contiguous negative and positive bins to determine the region of 
variable components (Paper I).
We mark all pixels where the difference is larger than 5\% of the average value and more 
than 3$\sigma$. Adjacent marked pixels 
are then connected to form a variable region. Then we expand such regime into neighboring 
pixels that have the same sign in the difference spectrum but at the less than 3$\sigma$ significant 
level. After that, we merge the neighboring regions with the same variable sign 
and with a separation of less than four pixels. Then it will generate the variable region
of BAL.

After the visual check, %there are 3116 pairs in 985 quasars with detected variable absorption lines were identified.
a sample containing 2324 spectrum pairs from 843 quasars with detected variable absorption lines (\CIV) and with 
the variations of continuum (1500\AA) beyond 5\% is identified (Fig. 2).
%We list the sources and regimes of the variable absorption lines in table \ref{tab1}. 

\begin{figure}
\center{}
\includegraphics[height=8.cm]{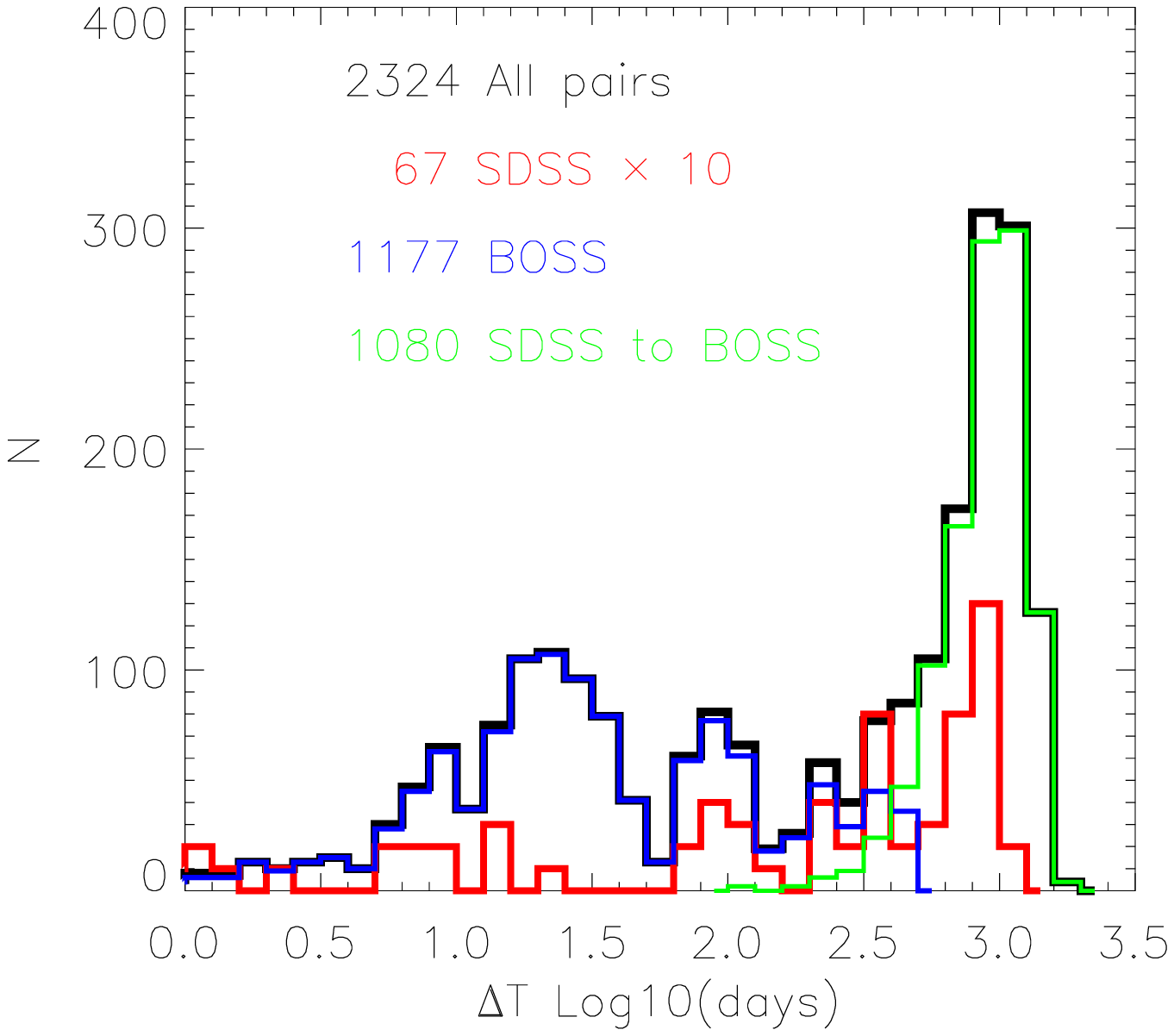}
\caption{The distributions of the time interval between two observations in the sample. 
There are 2324 spectrum pairs in 843 quasars with detected variable absorption lines and the 
variations of continuum beyond 5\% from the SDSS DR12. There are 67 pairs from the 
SDSS/only (MJD $\leq$ 55176) and 1177 pairs from the BOSS/only (MJD $>$ 55176). }
\end{figure}

\section{an indicator of the ionization state}
\label{sec:indicator}

\begin{figure}
\center{}
\includegraphics[height=10.1cm]{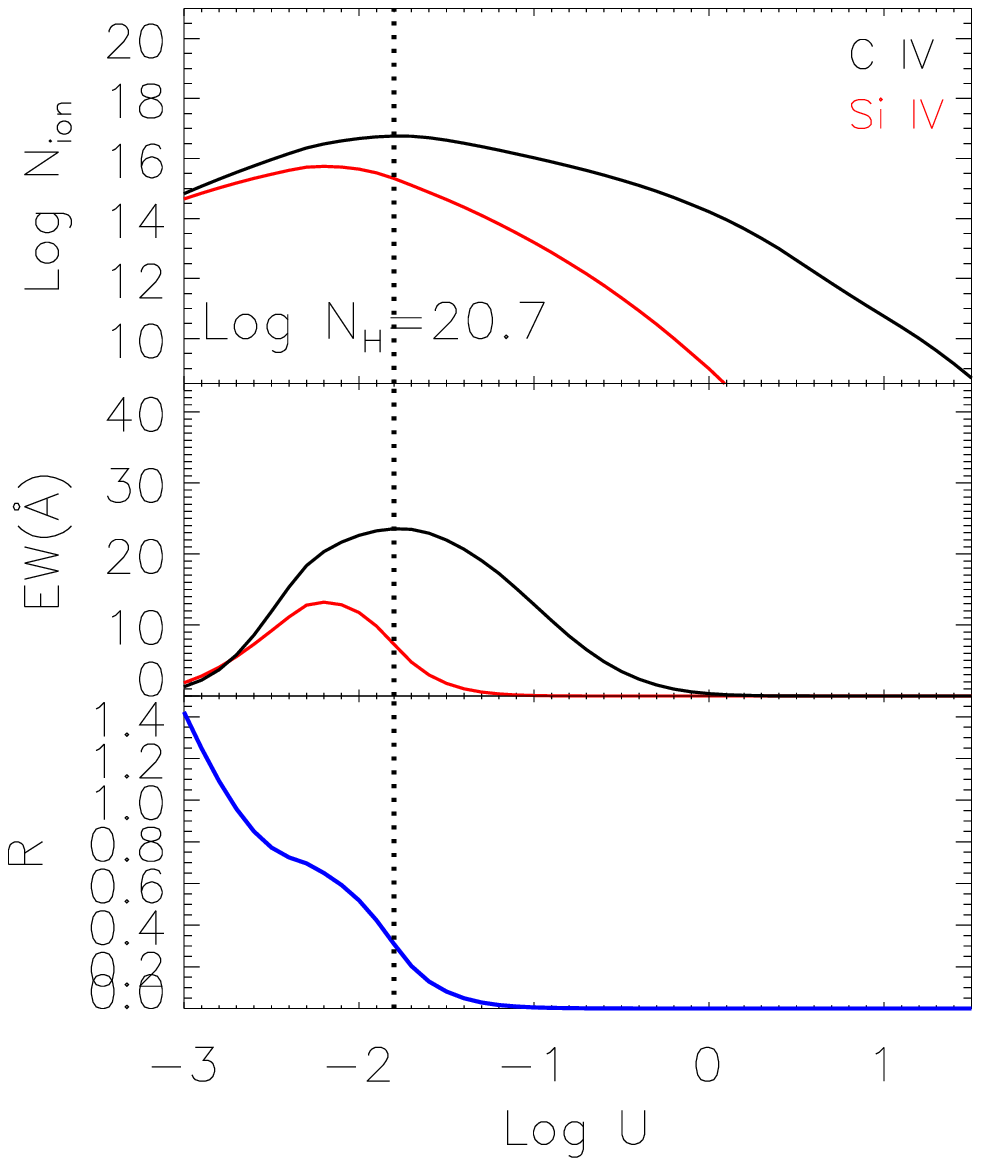}
\caption{Top panle: column densities respond to the variations of ionizing 
continuum using version c13.03 of Cloudy, described by Ferland et al. (2013).
The \civ~and \siiv~column densities increase first, then reach a peak, and decrease after 
as the ionization parameter increases, by the photoionization model using UV-soft SED ionizing continuum shape 
previously used for high-luminosity, radio-quiet quasars (Dunn et al. 2010). 
The hydrogen column density is $N_{\rm H}=10^{20.7}\cmii$. The vertical lines mark the $Log \rm U$ 
at the maximum column density of \civ. Second panle: the corresponding EWs of \siiv~and \civ~curves assuming a 
gaussian (FWHM=$2350\kms$) distribution of optical depth. Bottom panle: the EW ratio (R) of \siiv~to \civ~BAL 
decreases as the ionization parameter increases.}
\end{figure}
We run a series of photoionization simulations to see how ionic column densities respond to the variations of ionizing 
continuum using version c13.03 of Cloudy, described by Ferland et al. (2013). We consider a typical gas density of $10^6\cc$, 
since gas ionization is not sensitive to density at a given ionization parameter ($U=Q_{\rm H}/[4\pi r^{2}n_{\rm H}c] $,
where $Q_{\rm H}$ is the source emission rate of hydrogen ionizing photons, r is the distance to the absorber from 
the source, c is the speed of light, and $n_{\rm H}$ is the hydrogen number density).
As demonstrated in Fig. 3, the \civ~and \siiv~column densities increase first, then reach a peak, and decrease after 
as the ionization parameter increases. 
Due to the lower ionization potential, \siiv~will reach the peak earlier 
than \civ~as the ionization parameter increases. Assuming a gaussian distribution of optical depth, the EWs of
\siiv~and \civ~BALs can be deduced. The EW ratio of \siiv~to \civ~BALs:
\begin{equation}
\rm R_{\rm EW}=\frac{\rm EW_{\rm \siiv}}{\rm EW_{\rm \civ}},
\label{eq2}
\end{equation}
decreases as the ionization parameter increases at a given $N_{\rm H}$.
Generally speaking, $C^{+3}$ column density increases (decreases) if there is a (or no) $C^{+3}$ 
ionization front while the R decreases with incident continuum.
The R values should have the significant statistical differences between the high and low ionization 
states.%, although it can't give the exact value of ionization parameter.

For each of the spectrum pairs, we take the mean value of the two observations $R=(R_{1}+R_{2})/2$ as the indicator of 
the ionization state.
%, while the error of R is $\sigma_{\rm R}=\sqrt{{\sigma_{\rm R_{1}}}^2+{\sigma_{\rm R_{2}}}^2}/2$.

\section{The distributions of R for T1 and T2}
\label{sec:distribution}
As demonstrated in Fig. 3, the ionic column density of a specific species 
may respond to a continuum variation, negatively (EW of the BAL decreases when the continuum brightens) 
or positively, depending on the ionization of absorbing gas.
We divide the sample into two groups according to the pattern of the variation of BAL with respect to 
that of continuum: the EW of the BAL decreases (increases) when the continuum 
brightens (dims) as group T1; and the variation of EW and continuum in the opposite relation as group T2. 
As a result, the distributions of R for T1 and T2 should be different, as long as the BAL variabilities are driven by 
changes of the ionizing continuum. Further more, the mean value of R distribution for T1 should smaller than 
that for T2 in statistics.
\subsection{Using the intrinsic Baldwin effect} 
In paper I, we used the variation of emission-line EW as an independent check for the continuum variation.
72.5$\pm$2.2\% of EWs of the \civ~emission lines decrease (increase) when the continuum brightens (dims), 
which is 10.2$\sigma$ from no preference (50\%). The intrinsic Baldwin effect (e.g., Kinney et al. 1990) is 
detected at a higher conficence level (23.2$\sigma$) in this paper. So, we firstly use the variation of emission 
lines instead of the continuum variation to check the distributions of R for T1 and T2. We check both the mean
value (Student's T Test, T) and distributions of R (Kolmogorov-Smirnov test, KS). 
As demonstrated in Fig. 4, the mean value of R distribution for T2 is significantly ($P_{\rm T}=1.7E-02$) larger than that for T1. 
And the distributions of R for T1 and T2 are significantly different ($P_{\rm KS}=3.9E-06$).
It is the same as the photoionization model predictions. Like Paper I, about 75\% of the spectrum pairs are at
the high ionization state. 

\begin{figure}
\center{}
\includegraphics[height=6.4cm]{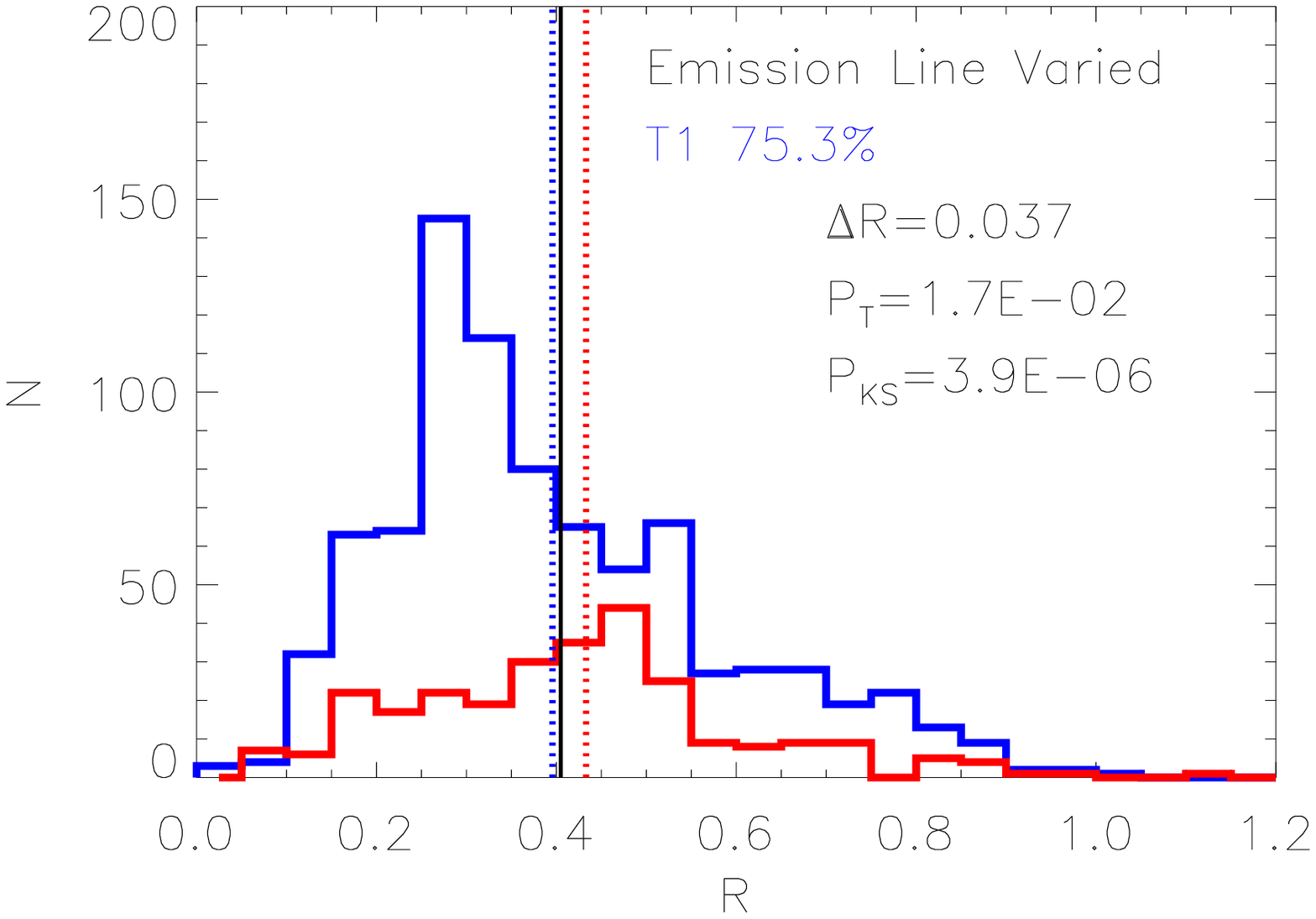}
\caption{The distributions of R for T1 and T2 using the variation of emission lines. Blue line is for T1 while
the red one is for the T2. The mean value of R distribution for T2 (red vertical dashed line) 
is significantly ($P_{\rm T}=1.7E-02$) 
larger than that for T1 (blue one). And the distributions of R for T1 and T2 are significantly 
different ($P_{\rm KS}=3.9E-06$).
The black solid vertical line represents the mean value of R distribution with T1 and T2 combined.}
\end{figure}

\subsection{Sub-sample of SDSS and BOSS} 
As shown in Fig. 2, there are total 67 spectrum pairs from the SDSS/only (MJD $\leq$ 55176) and 1177 
pairs from the BOSS/only (MJD $>$ 55176).
Like Paper I, we use the variability amplitude $>$5\% of continuum flux as a threshold.
It is shown in Fig. 5 that the mean value of R distribution for T2 is significantly larger than that for
T1 ($\Delta R$ = 0.164, $P_{\rm T}=1.2E-02$ for SDSS; $\Delta R$ = 0.033, $P_{\rm T}=2.6E-04$ for BOSS). 
The distributions of R for T1 and T2 are significantly different for both SDSS ($P_{\rm KS}=2.5E-03$) and BOSS ($P_{\rm KS}=1.1E-12$).
It is also the same as predicted by the photoionization model. The $\Delta R$ of the SDSS is larger than that of the BOSS. 
The reason maybe that the calibration uncertainties of BOSS is larger than that of the SDSS. 

\begin{figure}
\center{}
\includegraphics[height=6.4cm]{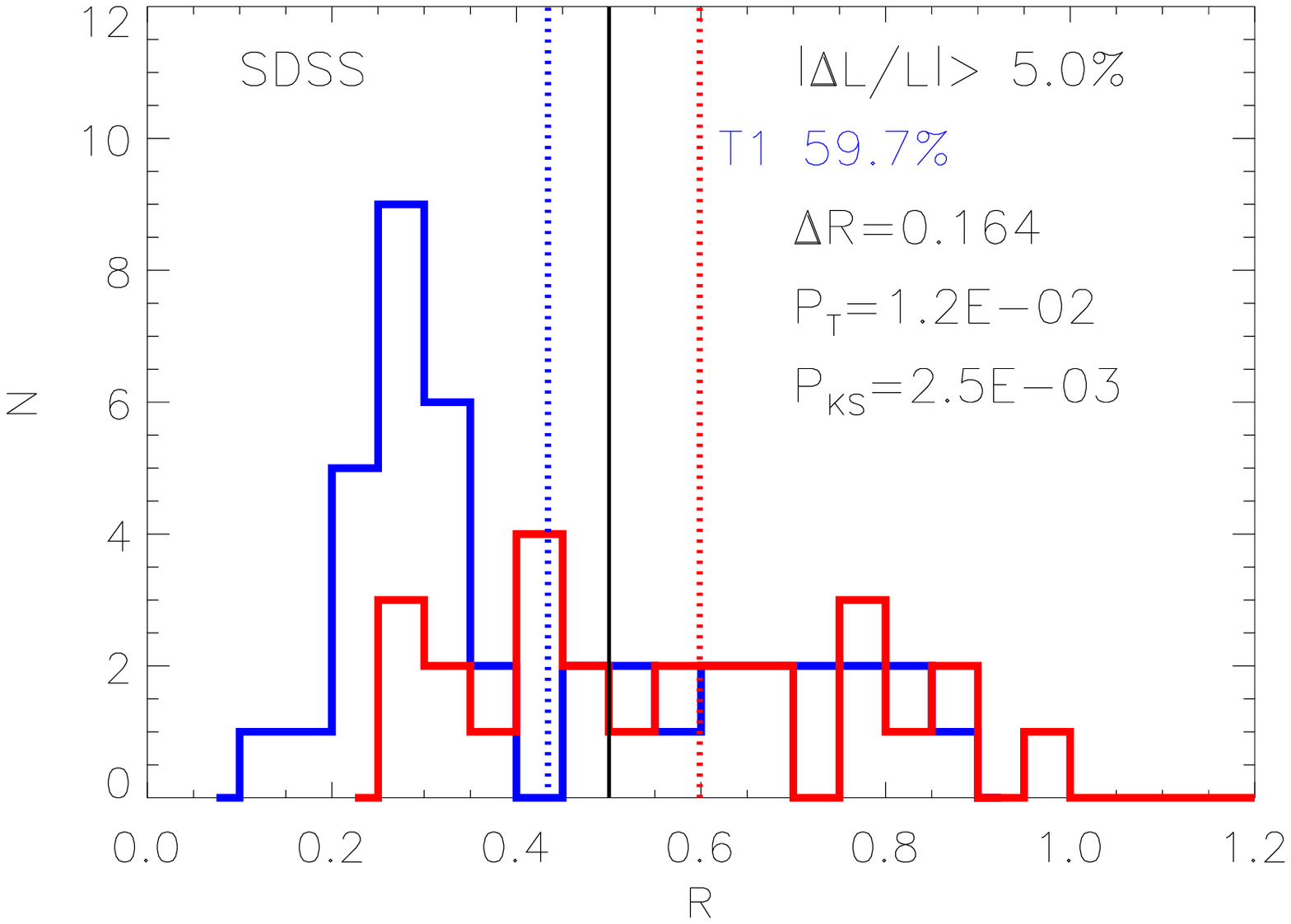}
\includegraphics[height=6.4cm]{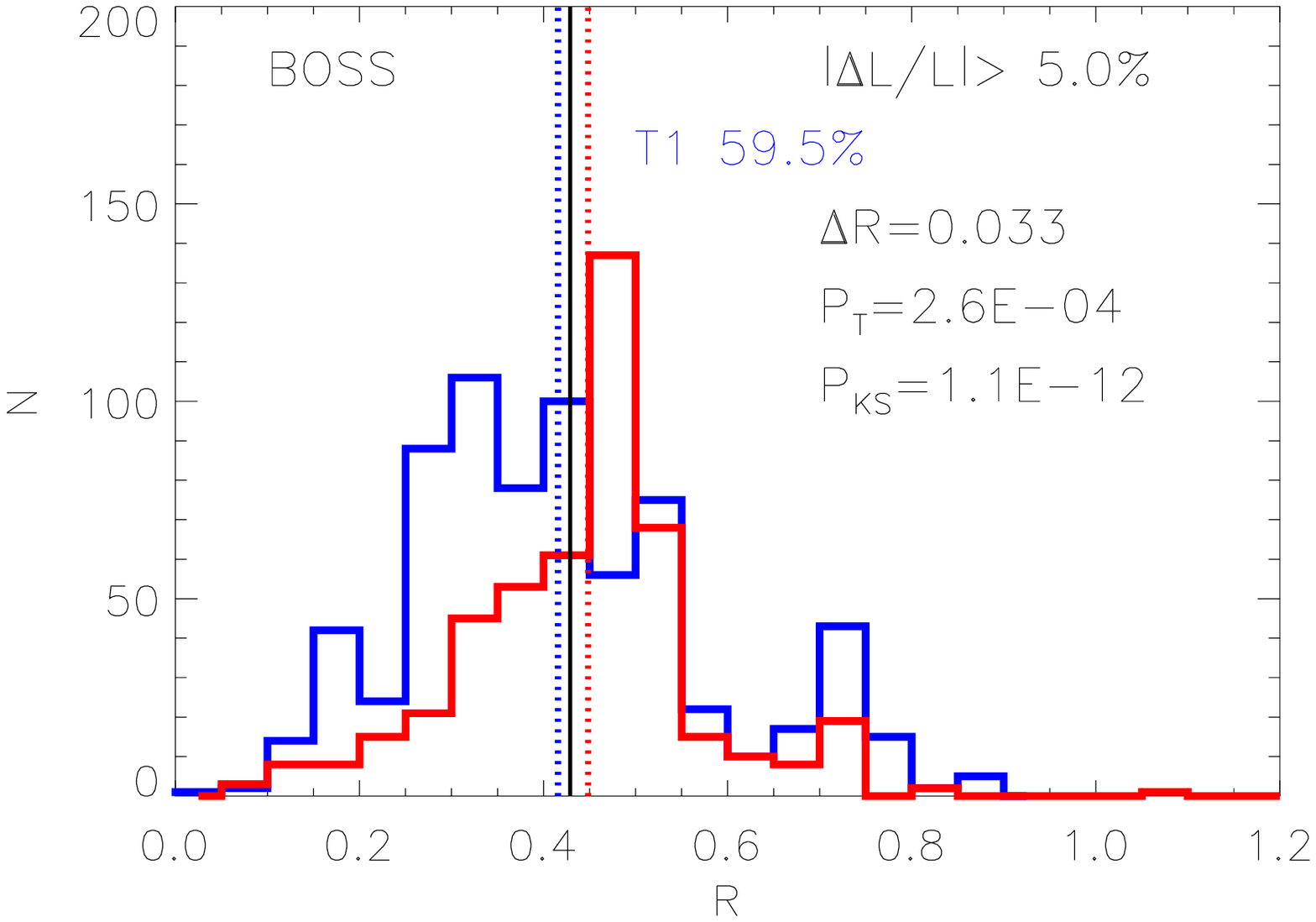}
\caption{The distributions of R for T1 and T2. Top is for the SDSS and bottom is for the BOSS.
The mean value of R distribution for T2 is significantly
larger than that for T1 for both SDSS ($P_{\rm T}=1.2E-02$) and BOSS ($P_{\rm T}=2.6E-04$).
And the distributions of R for T1 and T2 are significantly different for both SDSS ($P_{\rm KS}=2.5E-03$)
and BOSS ($P_{\rm KS}=1.1E-12$). The symbols are the same as in Fig. 4.}
\end{figure}

\subsection{The whole sample} 
There are total 2324 spectrum pairs from 843 quasars with detected variable absorption lines and the 
variations of continuum beyond 5\%. We gradually increase the amplitude of 
variations of continuum to compare the distributions of R for T1 and T2. As shown in Fig. 6,
the distributions of R for T1 and T2 are significantly ($P_{\rm KS}=5.4E-19$) different. 
The mean value of R distribution for T2 is significantly ($P_{\rm T}=7.5E-07$) larger than that for T1 
and the $\Delta R$ increases with the amplitude of variations 
of continuum. %As mentioned above, about 75\% of the spectrum pairs are at the high ionization state (T1). 
%{\color{blue}The ionic potential of \aliii~is 28.4ev which is lower than the \civ~and \siiv. 

Above distributions of R for T1 and T2 are all significantly different. And the mean value 
of R distribution for T2 is significantly larger than that for T1. It agrees with the predication 
of the photoionization model. These evidences support the idea that variation of the ionizing continuum 
is the main driver for BAL variability.

\begin{figure*}
\center{}
\includegraphics[height=6.3cm]{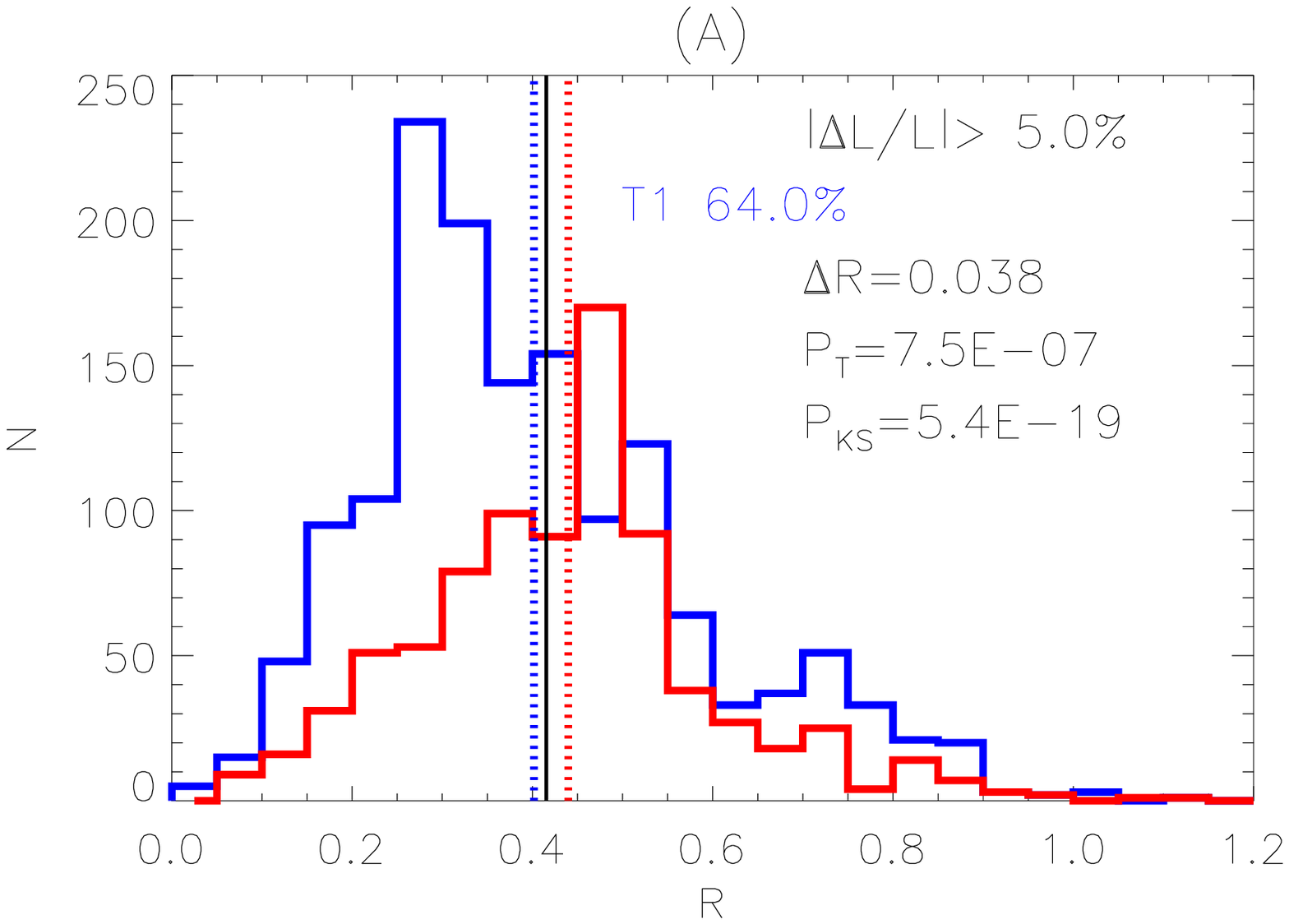}
\includegraphics[height=6.3cm]{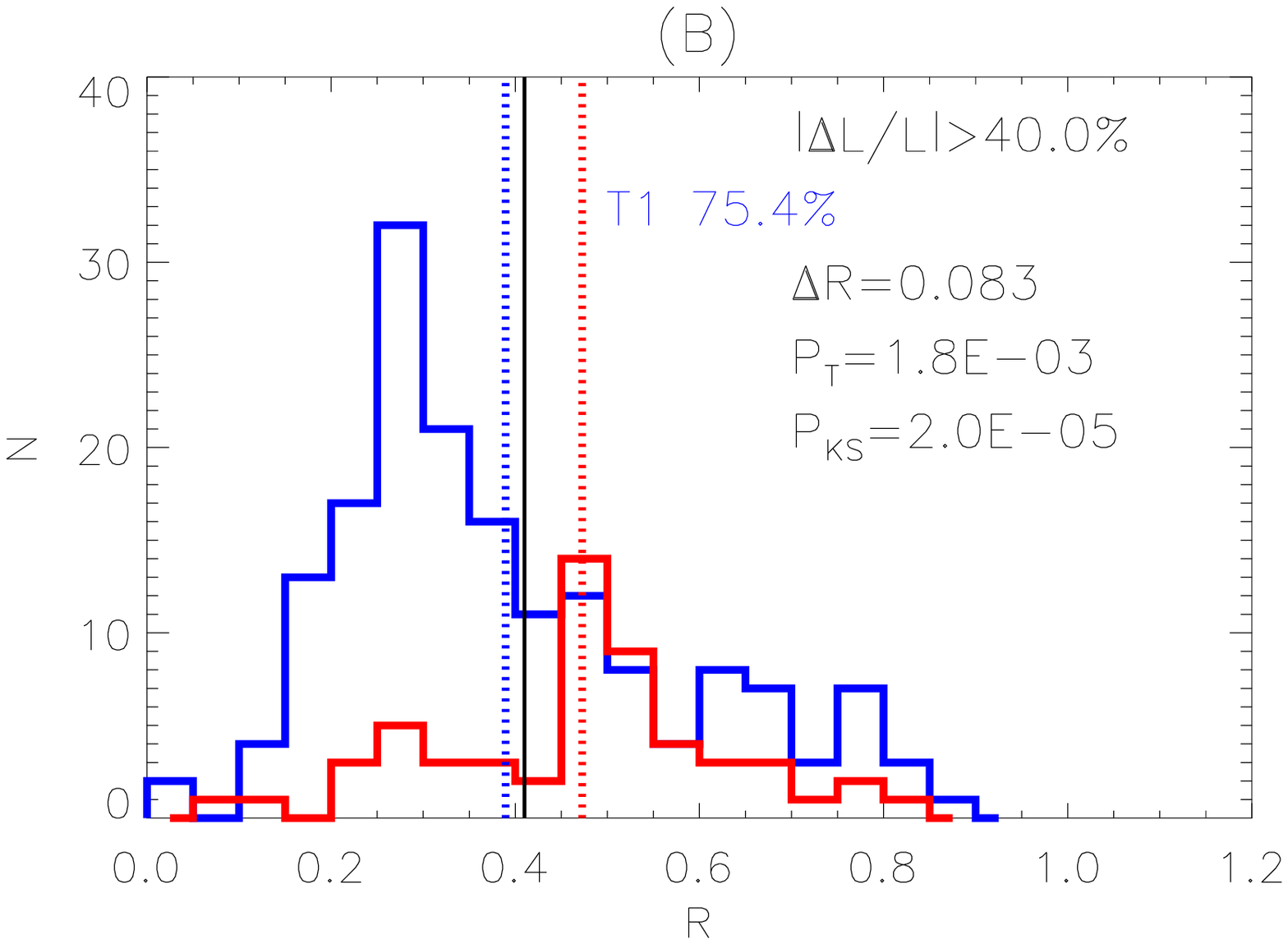}
\includegraphics[height=13.5cm]{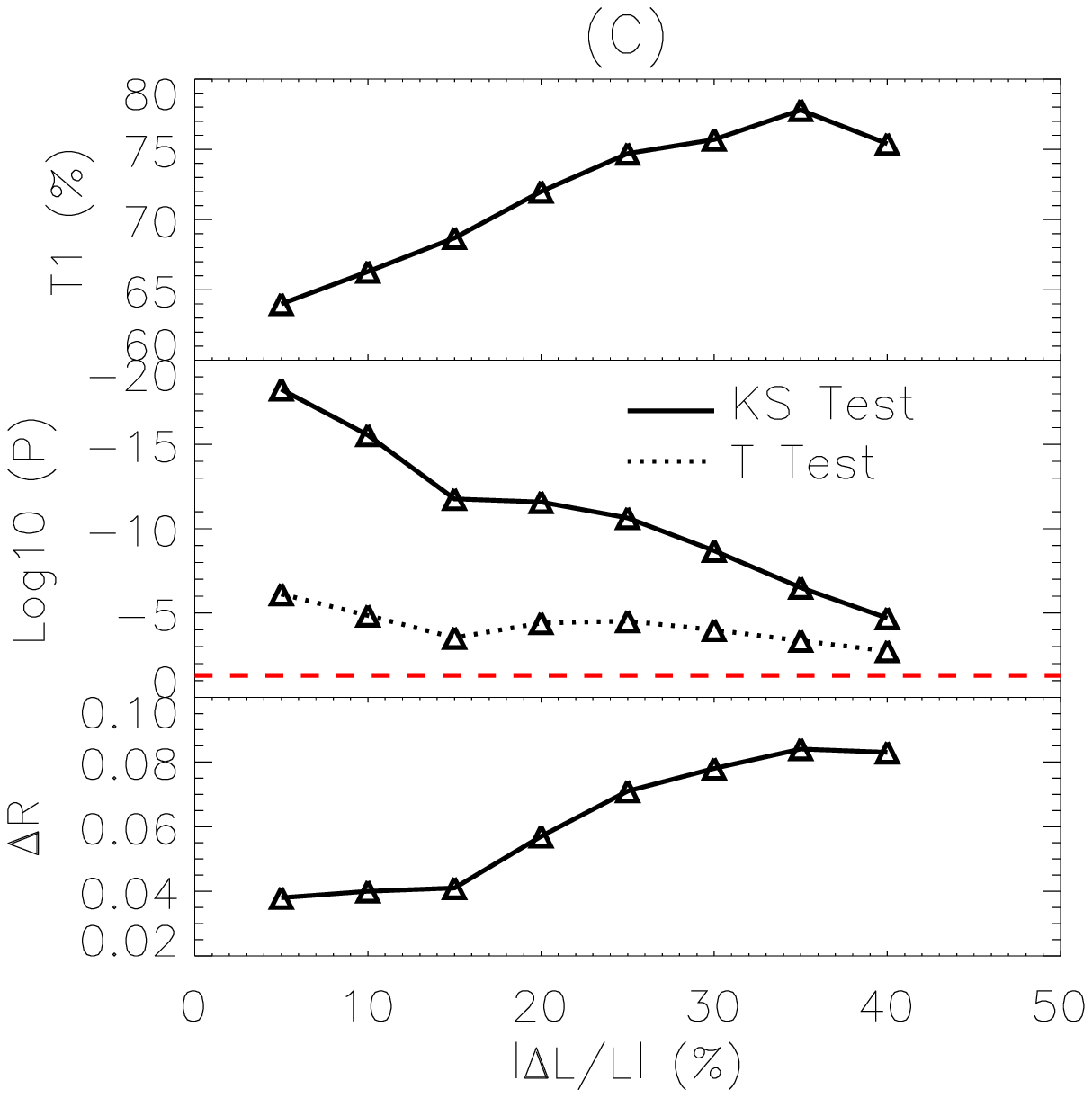}
\caption{(A) and (B): the distributions of R for T1 and T2 at $|\Delta L/L|>5\%$ and $40\%$ . 
The mean value of R distribution of T2 is significantly larger than the T1.
And the distribution of R for T1 and T2 is significantly different. The symbols are the same as in Fig. 4.
(C): the fraction of T1 (top panle), p vaule (middle panle) of KS Test (solid line) and T Test (dotted line), 
$\Delta R $ (bottom panle), for the distributions of T1 and T2 at different thresholds of continuum variability.
The red dotted horizontal line marks the $p=0.05$ confidence level. The fraction of T1 and $\Delta R $ gradually increase
with the thresholds of continuum variability.}
\end{figure*}

\subsection{Estimate of the fraction of BAL variability driven by the photoionization } 
On the one hand, the uncertainty of flux calibration doesn't affect the measurement of BAL EW.
On the other hand, the variation of covering factor of the BAL gas doesn't change the UV continuum.
As a result, considering the uncertainty of flux calibration or the variation of covering factor of the outflow,
part of the BALs will be randomly assigned to the two types (T1 and T2). The BALs which originally belong 
to T1 (T2), may be assigned to T2 (T1) at the same probability ($P_{mis}$ in Fig. 7) for the two types.
As shown in top panle of Fig. 7, the fraction of observed T1 is A. The fraction of misassignment from T1 to T2 is n
, while the fraction from T2 to T1 is m. Then the misassignment possibility is $P_{mis}=n/(A-m+n)=m/(1-A-n+m)$.
Then, the fraction of BAL variability driven by the photoionization is $F_{pho} = 1-2(n+m)$.

We use the skew gaussian function,
\begin{equation}
N(r)=\large a_{1}\exp \left[\frac{-(r-a_{2})^2} {2a_{3}^2}\right]\left( 1+erf[a_{4}(r-a_{2})]\right)
%N(r)=a\times e ^{-\left(\frac{r-b} {c}\right)^2 } \times \left( 1+erf[d\times (r-b)~] ~\right)
\label{eq3}
\end{equation}
to describe the intrinsic distributions of T1 and T2 of our sample, 
where $erf(r)=\int_{-r}^{r}e^{-t^2} dt$ is the error function and the parameter $a_{4}$ is the coefficient of skewness. 
The observed T1 consists of two components:
the blue dotted line and the red shadow region while T2 consists of the red dotted line and the blue shadow region.
The blue (red) shadow region is the fraction of misassignment from T1 (T2) to T2 (T1). We fit the distributions 
combining sub-samples at different thresholds of continuum variability. During the fitting, we tie all the parameters
together except $a_{1}$ which varies independently for sub-samples at different thresholds of continuum variability,
with the minimizing $\chi^2=2.29$.
The error of each bin of the distributions is estimated using the bootstrap method. 

It shows that $F_{pho}$ increases with the thresholds of continuum variability. This is expected because the 
fraction of misassignment caused by the uncertainty of flux calibration gradually decreases with 
the amplitude of continuum variability. $F_{pho}$ is about 80\% when the thresholds of continuum variability 
is beyond 30\%. As shown in Fig. 8, the R isn't thoroughly monotonic with ionization parameter and there exists 
a plateau when the $N_{\rm H}$ is large. The plateau elongates as the $N_{\rm H}$ increases. The intrinsic distribution 
of T1 is broader than T2 from the fitting results. This may imply that many T2 quasars are at the plateau.
%And quite a lot of BAL outflows have a relatively high $N_{\rm H}$.
\begin{figure*}
\center{}
\includegraphics[height=7.cm]{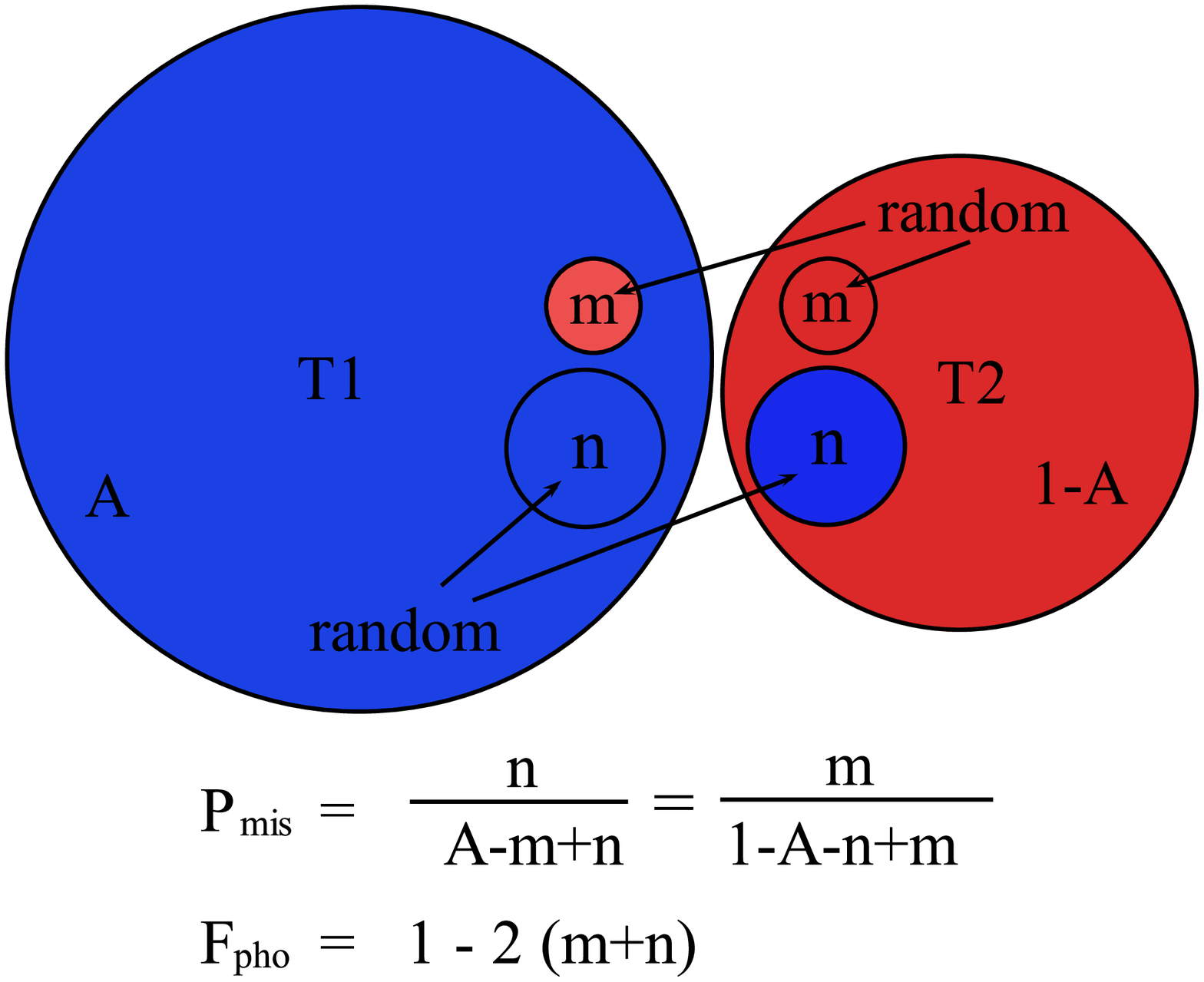}
\includegraphics[height=13.cm]{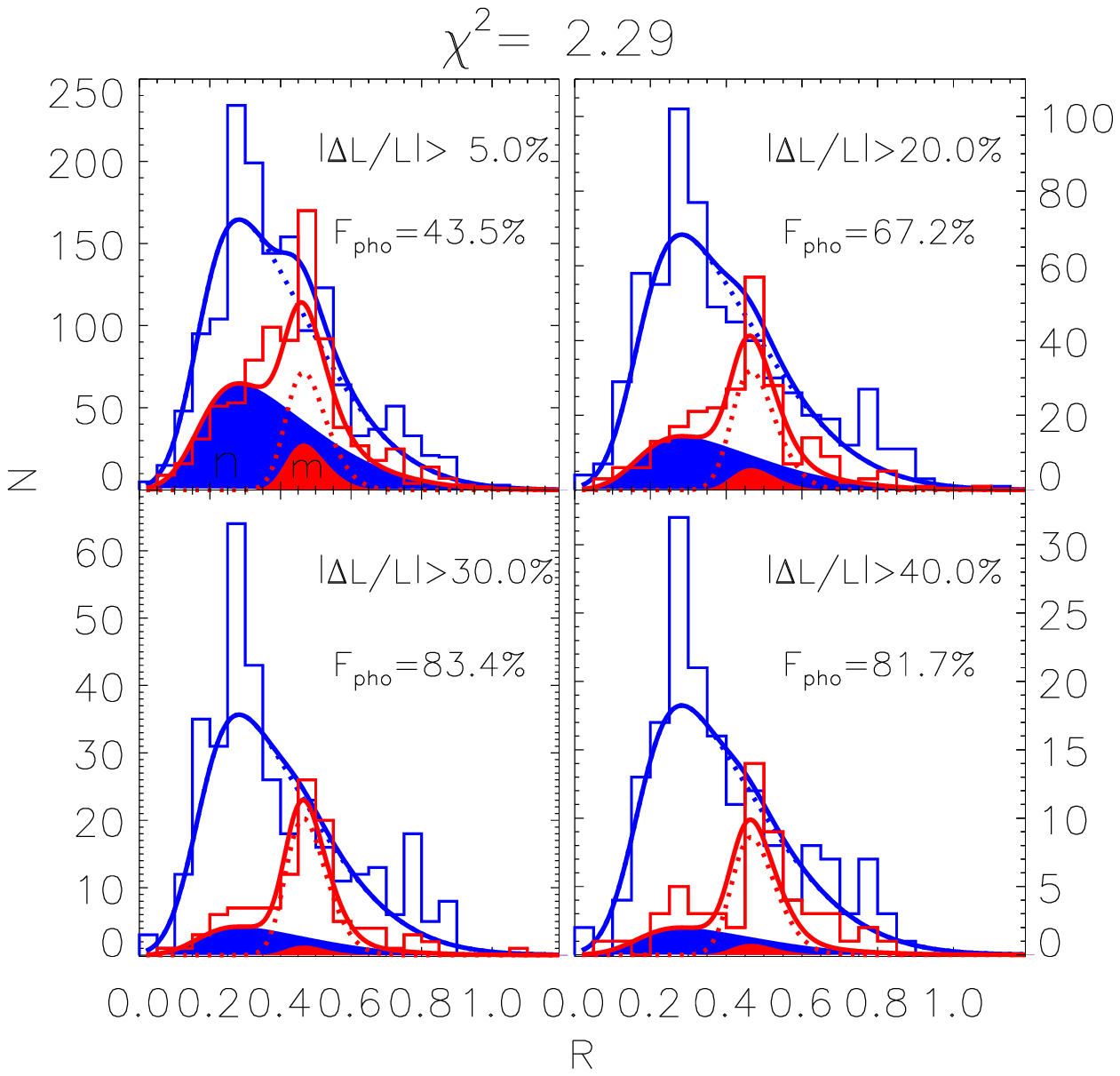}
\caption{Top panle: assuming that the BALs which originally belong to T1 (T2), may be assigned to T2 (T1) at the same 
probability. The fraction of observed T1 is A. The fraction of misassignment from T1 to T2 is n, 
while the fraction from T2 to T1 is m. Then the misassignment possibility is $P_{mis}=n/(A-m+n)=m/(1-A-n+m)$.
The fraction of BAL variability driven by the photoionization is $F_{pho} = 1-2(n+m)$.
Bottom panle: using the skew gaussian function to fit the intrinsic distributions of T1 and T2. 
T1 consists of two components:
the blue dotted line and the red shadow region while T2 constituts of the red dotted line and the blue shadow region.
The blue (red) shadow region is the fraction of misassignment from T1 (T2) to T2 (T1).
$F_{pho}$ gradually increases with the thresholds of continuum variability. $F_{pho}$ is about 80\% when the 
threshold of continuum variability is beyond 30\%.}
\end{figure*}

\begin{figure} 
\center{}
\includegraphics[height=10.cm]{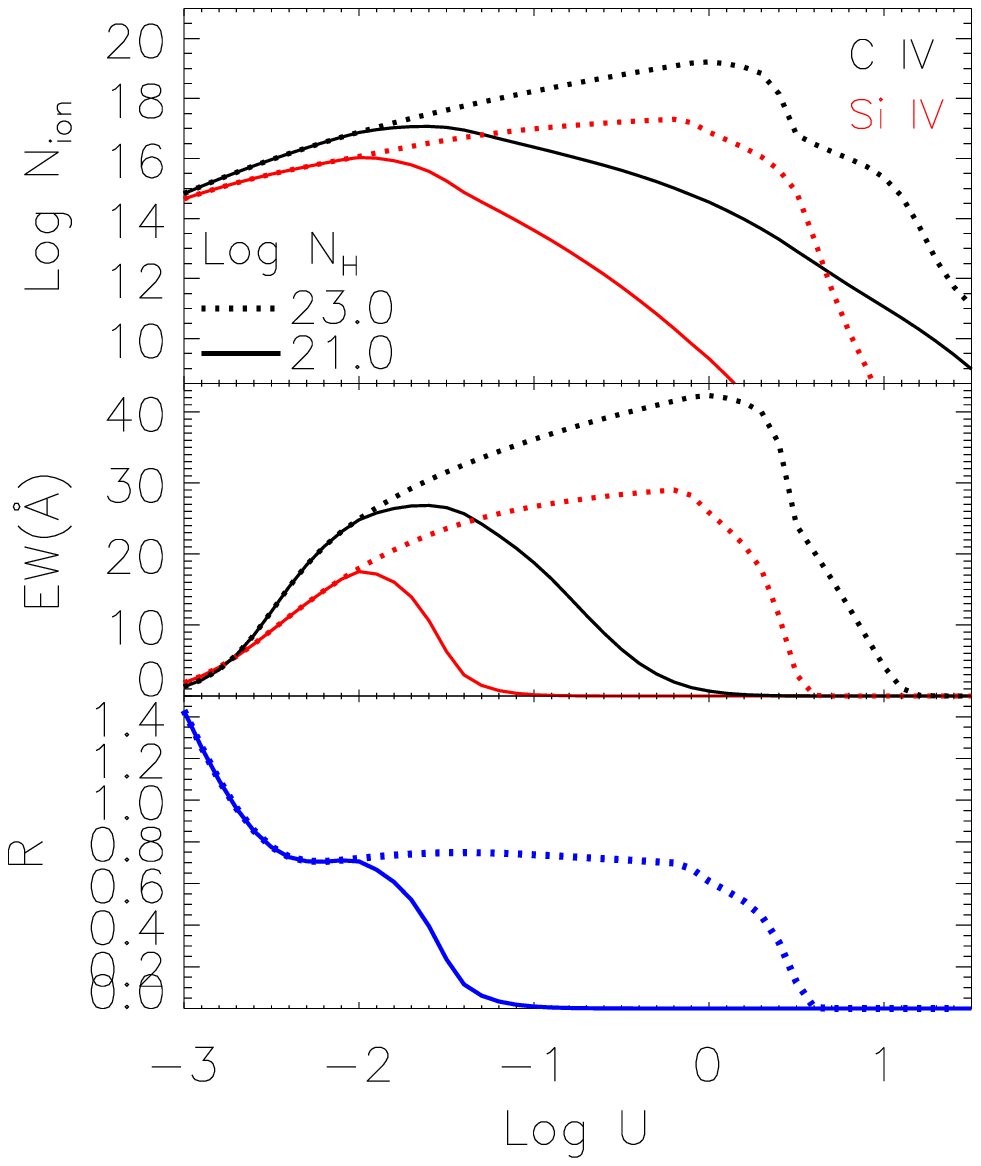}
\caption{The EW ratio (R) of \siiv~to \civ~decreases as the ionization parameter increases while there exists a plateau. 
The plateau elongates as the $N_{\rm H}$ increases. Solid line is the gas column density $N_{\rm H}=10^{21}\cmii$ 
while dotted line is $N_{\rm H}=10^{23}\cmii$.}
\end{figure}

\section{The recombination time scale} 
\label{sec:timescale}
As mentioned in Paper I, the high concordance between continuum and absorption line variations
requires a recombination time $t_{rec} = (\alpha n_{e})^{-1}$ shorter than both the time scale of typical 
continuum variations and the interval between the two observations (e.g., Barlow et al. 1992), where $\alpha$
is the recombination rate and $n_{e}$ is the electron density .
As a result, if the time scale of typical continuum variations or the interval between the two-epochs of the spectrum pair
is shorter than the recombination time, there will be no significant difference between the distributions 
of R for T1 and T2. We gradually change the time interval between the two-epochs and the amplitude of 
variations of continuum. As shown in Fig. 9, it still has a significant difference ($P_{\rm KS}=0.031$, $P_{\rm T}=0.032$) 
between the distributions of T1 and T2 when $\Delta T < 6~days$ and the amplitude of variations of 
continuum is larger than 15\%. 
%There has no significant difference between the distributions of R when the $\Delta T < 5~days$. 
This result suggests that there exists the BAL outflows whose recombination time scales
are only about several days. 
Here note that, $\Delta R$ clearly increases (0.019 to 0.134) with the amplitude of 
variations of continuum (5\% to 25\%) when $\Delta T <5~days$. It implies that the upper limit of
recombination time scale may be shorter than $5~days$. 
It agrees with the result of Grier et al. 2015 which they found the shortest BAL
variability time scale (1.2days) yet reported. They concluded that it's most likely a rapid response to 
changes in the ionizing continuum.
%Due to the small size of the sample at $\Delta T <5~days$, there is no
%significant statistical difference between the T1 and T2. 

For low-redshift AGN, the most direct method to determine the outflow radius is using spatially resolved spectroscopy,
especially integral field unit (IFU) spectroscopy, e.g. (Barbosa et al. 2009; Riffel \& Storchi-Bergmann 2011;
Rupke \& Veilleux 2013; Liu et al. 2013a,b, 2014, 2015), but for high-redshift ($z>2$) BAL quasars,
the realistic approach is deriving the galactocentric distance of
the outflow by the hydrogen number density $n_{\rm H}$ which can be determined
from the absorption lines of the excited states of ions (e.g. Fe II*, Si II*, 
S IV*). During the last decade or so, the outflow radii have been measured
for a number of individual quasars using density-sensitive absorption lines
from excited levels (Arav et al. 1999; de Kool et al. 2001, 2002a, b;
Hamann et al. 2001; Arav et al. 2008; Moe et al. 2009; Bautista et al. 2010; Dunn et al. 2010; Aoki et al. 2011;
Borguet et al. 2012; Arav et al. 2013, 2015; Chamberlain et al. 2015).
Due to line blending, the method can only apply to relatively narrow line. 
Complementary to these conventional approaches, we can monitor some quasars continuously whose BAL variabilities
and recombination time scales are only several days. 
Then we can constrain on the electron density from the time lags of BAL to continuum variability.

\begin{figure} 
\center{}
\includegraphics[height=7.5cm]{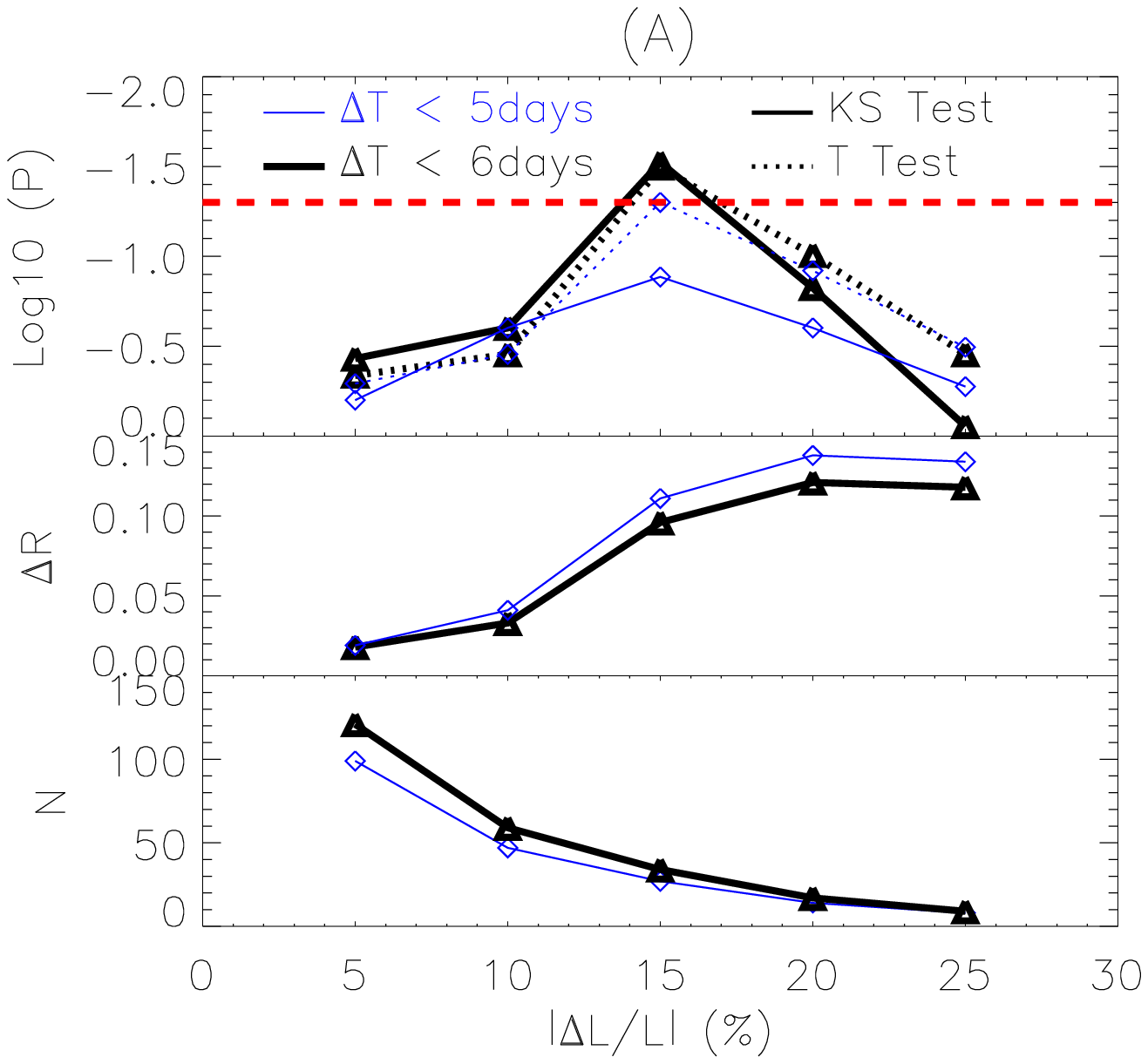}
\includegraphics[height=6.3cm]{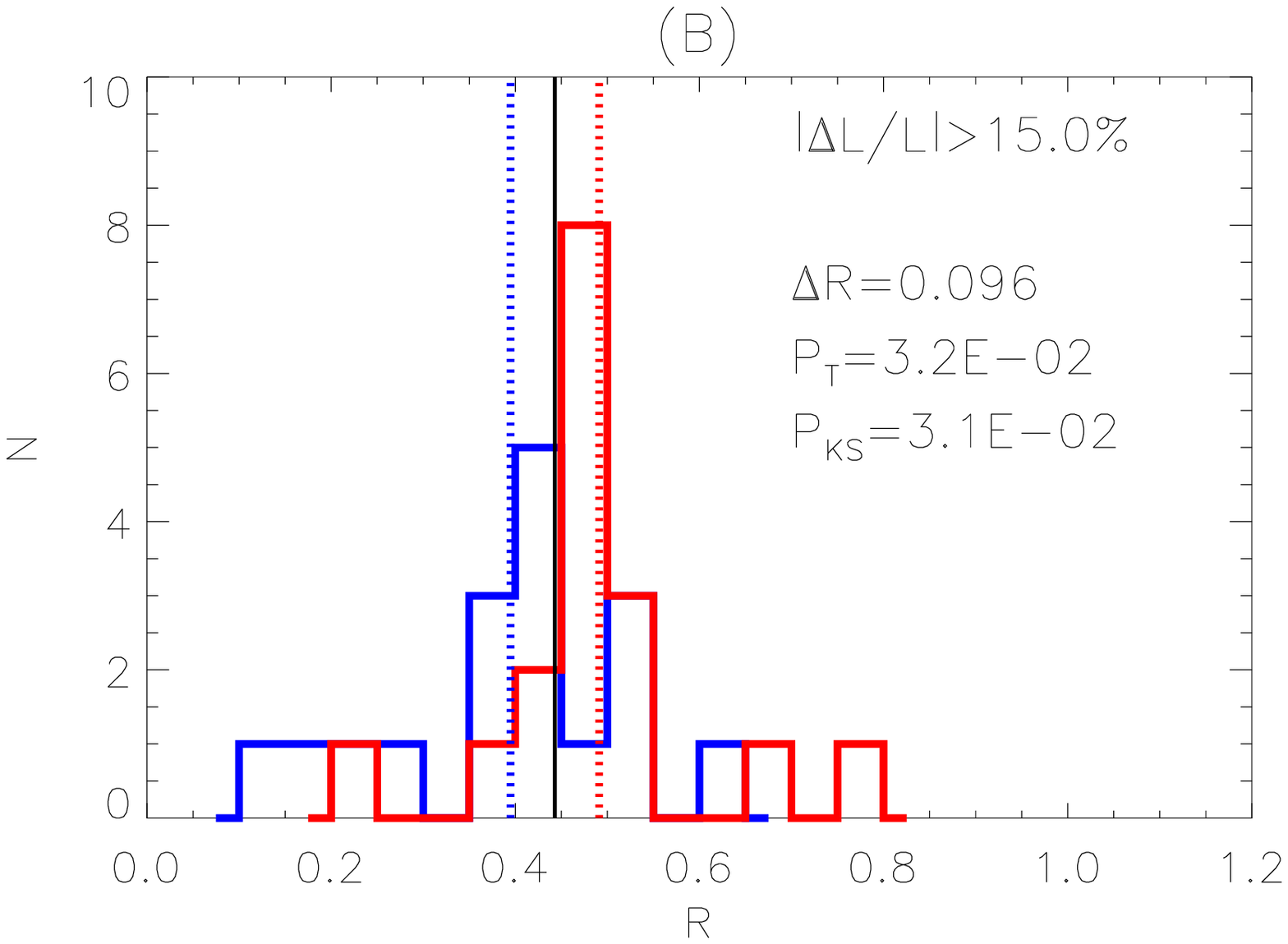}
\caption{Using the distributions of R for T1 and T2 to constrain the upper limit of the recombination time scale.
(A): the p vaule (top panle) of KS Test (solid line) and T Test (dotted line), $\Delta R $ (middle panle), 
sample size (bottom panle) for the distributions of T1 and T2 at different thresholds of continuum variability.
Black line is $\Delta T <6~days$, while blue one is $\Delta T <5~days$. The red horizontal dotted line marks the $p=0.05$ 
confidence level. (B): both the p vaules of KS and T Test are lower than 0.05 with $|\Delta L/L|>15\%$ and $\Delta T <6~days$.
As a result, a upper limit of the tpyical recombination time scale about 6 days is given by our sample.
The symbols are the same as in Fig. 4.}
\end{figure}

\section{Conclusions}
\label{Conclusions}
In this paper, we present a statistical analysis of the variability of BAL for a sample of 2324 spectrum 
pairs in 843 quasars (1.9 $<$ z $<$ 4.7) with detected variable absorption lines and the variations of
continuum beyond 5\% from the SDSS DR12. The main results are as follows:

(1) Our study shows that the distributions of R for T1 (high ionization state) and T2 (low ionization state) 
are significantly ($P_{\rm KS}=5.4E-19$) different, 
while the mean value of R of T1 is significantly ($P_{\rm T}=7.5E-07$) lower than the T2. 
These results suggest that the variabilities of BALs are dominated by the variation of ionizing continuum.

(2) Considering the uncertainty of flux calibration and the variation of covering factor of 
the outflow, part of the BALs will be randomly assigned to the two types. As a result, the significance of 
difference of R for T1 and T2 will be weakened by above two aspects.
We give an estimate for the fraction of BAL variability driven by the variation of ionizing continuum. The fraction of 
BALs variability driven by the variation of ionizing continuum increases with the thresholds of continuum variability.
It is about 80\% when the thresholds of continuum variability is beyond 30\%.

(3) It still has a significant difference ($P_{\rm KS}=0.031$, $P_{\rm T}=0.032$) 
between the distributions of T1 and T2 when the $\Delta T <$ 6 days and the amplitude of variations of 
continuum is larger than 15\%. This result suggests that there exists the BAL outflows whose recombination time scales
are only about several days. To determine the electron density, we can monitor some quasars continuously 
whose BAL variabilities and recombination time scales are short.
Once the density is known, the size of the BAL outflow can be measured.

\section{ACKNOWLEDGMENTS}
Authors thank the anonymous referee for constructive
comments and suggestions for improving the clarity of the
manuscript.
We acknowledge the financial support by the Strategic Priority Research Program "The Emergence of Cosmological 
Structures" of the Chinese Academy of Sciences (XDB09000000), NSFC (NSFC-11233002, NSFC-11421303, U1431229),
National Basic Research Program of China (grant No. 2015CB857005) and National Science Foundation of China
(nos. 11373024 \& 11233003).

Guilin Liu is supported by the National Thousand Young Talents Program of China, and acknowledges the grant from the National Natural Science Foundation of China (No. 11673020 and No. 11421303) and the Ministry of Science and Technology of China (National Key Program for Science and Technology Research and Development, No. 2016YFA0400700).

%We thank the referee for useful comments. 
%We acknowledge the financial support 
%by the Strategic Priority Research.


\begin{thebibliography}

\bibitem[Allen et al.(2011)]{2011MNRAS.410.860} Allen, J. T., et al.\ 2011, \mnras, 410, 860
\bibitem[Aoki et al.(2011)]{2011PASJ.63.457}Aoki, K., Oyabu, S., Dunn, J. P., Arav, N., Edmonds, D.,
Korista, K. T., Matsuhara, H., Toba, Y., 2011, PASJ, 63, 457
\bibitem[Arav et al.(1999)]{1999AAS.195.1805}Arav, N., de Kool, M., Becker, R. H., Laurent-Muehleisen, S. A., White, R. L., Price, T., Gregg, M. D., 1999, AAS, 195, 1805 
\bibitem[Arav et al.(2008)]{2008ApJ.681.954}Arav, N., Moe, M., Costantini, E., Korista, K. T., Benn, C.,
Ellison, S., 2008, ApJ, 681, 954
\bibitem[Arav et al.(2013)]{2013MNRAS.436.3286A} Arav, N., Borguet, B., 
Chamberlain, C., Edmonds, D., \& Danforth, C.\ 2013, \mnras, 436, 3286 
\bibitem[Arav et al.(2015)]{2015AA.577.37} Arav, N. et al., 2015, A\&A, 577, 37
\bibitem[Barbosa et al.(2009)]{2009MNRAS...396...2}Barbosa, F. K. B., Storchi-Bergmann, T., Cid Fernandes, R.,
Winge, C., \& Schmitt, H. 2009, MNRAS, 396, 2
\bibitem[Barlow et al.(1992)]{1992ApJ...397...81B} Barlow, T.~A., 
Junkkarinen, V.~T., Burbidge, E.~M., et al.\ 1992, \apj, 397, 81 
\bibitem[Bautista et al.(2010)]{2010ApJ...713...25}Bautista, M. A., Dunn, J. P., Arav, N., Korista, K. T., Moe, M., Benn, C., 2010, ApJ, 713, 25
\bibitem[Borguet et al.(2012)]{2012ApJ...751...107}Borguet, B. C. J., Edmonds, D., Arav, N., Dunn, J., Kriss,
G. A., 2012, ApJ, 751, 107
\bibitem[Capellupo et al.(2011)]{2011MNRAS.413..908C} Capellupo, D.~M., 
Hamann, F., Shields, J.~C., Rodr{\'{\i}}guez Hidalgo, P., 
\& Barlow, T.~A.\ 2011, \mnras, 413, 908 
\bibitem[Capellupo et al.(2012)]{2012MNRAS.422.3249C} Capellupo, D.~M., 
Hamann, F., Shields, J.~C., Rodr{\'{\i}}guez Hidalgo, P., 
\& Barlow, T.~A.\ 2012, \mnras, 422, 3249 
\bibitem[Capellupo et al.(2013)]{2013MNRAS.429.1872C} Capellupo, D.~M., 
Hamann, F., Shields, J.~C., Halpern, J.~P., 
\& Barlow, T.~A.\ 2013, \mnras, 429, 1872 
\bibitem[Capellupo, Hamann \& Barlow (2014)]{2014MNRAS.444.1893C} Capellupo, D.~M., 
Hamann, F., \& Barlow, T.~A.\ 2014, \mnras, 444, 1893  
\bibitem[Chamberlain et al.(2015)]{2015MNRAS.450.1085C} Chamberlain, C., Arav, N., Benn, C., 2015, \mnras, 450, 1085C
\bibitem[Dawson et al.(2013)]{2013AJ....145...10D} Dawson, K.~S., Schlegel, 
D.~J., Ahn, C.~P., et al.\ 2013, \aj, 145, 10 

\bibitem[de Kool et al.(2001)]{2001ApJ....548...609}de Kool, M, Arav, N, Becker, R H., Gregg, M D., White, R L., Laurent-Muehleisen, S A., Price, T, Korista, K T., 2001, ApJ, 548, 609
\bibitem[de Kool et al.(2002a)]{2002ApJ....567...58}de Kool, M., Becker, R. H., Gregg, M. D., White, R. L., Arav, N., 2002a, ApJ, 567, 58
\bibitem[de Kool et al.(2002b)]{2002ApJ....570...514}de Kool M., Becker R. H., Arav N., Gregg, M. D., White, R. L., 2002b, ApJ, 570, 514

\bibitem[Di Matteo et al.(2005)]{2005Nat...433.604}Di Matteo, T., Springel, V., Hernquist, L.\ 2005, Nat, 433, 604
\bibitem[Dunn et al.(2010)]{2010ApJ...709.611}Dunn, J. P. et al.\ 2010, \apj, 709, 611
%\bibitem[Emmering et al.(1992)]{1992ApJ...385.460}Emmering, R. T., Blandford, R. D., \& Shlosman, I.\ 1992, \apj, 385, 460
\bibitem[Ferland et al.(2013)]{2013....}Ferland, G. J., Porter, R. L., van Hoof, P. A. M., Williams, R. J. R., Abel, N. P. , Lykins, M. L., Shaw, G., Henney, W. J., \& Stancil, P. C.. The 2013 Release of Cloudy. Revista Mexicana de Astronomia y Astrofisica, 49:137163
\bibitem[Filiz Ak et al.(2012)]{2012ApJ...757..114F} Filiz Ak, N., Brandt, 
W.~N., Hall, P.~B., et al.\ 2012, \apj, 757, 114 
\bibitem[Filiz Ak et al.(2013)]{2013ApJ...777..168F} Filiz Ak, N., Brandt, 
W.~N., Hall, P.~B., et al.\ 2013, \apj, 777, 168 
%\bibitem[Filiz Ak et al.(2014)]{2014ApJ...791..88F} Filiz Ak, N., Brandt, 
%W.~N., Hall, P.~B., et al.\ 2014, \apj, 791, 88
\bibitem[Gibson et al.(2010)]{2010ApJ...713..220G} Gibson, R.~R., Brandt, 
W.~N., Gallagher, S.~C., Hewett, P.~C., \& Schneider, D.~P.\ 2010, \apj, 713, 220 
\bibitem[Gibson et al.(2009)]{2009ApJ...696..924G} Gibson, R.~R., Brandt, 
W.~N., Gallagher, S.~C., \& Schneider, D.~P.\ 2009, \apj, 696, 924 
\bibitem[Gibson et al.(2008)]{2008ApJ...675..985G} Gibson, R.~R., Brandt, 
W.~N., Schneider, D.~P., \& Gallagher, S.~C.\ 2008, \apj, 675, 985 
\bibitem[Grier et al.(2015)]{2015ApJ...806..111} Grier, C. J., et al.\ 2015, \apj, 806, 111
\bibitem[Hamann et al.(2001)]{2001ApJ.550.142} Hamann, F. W., Barlow, T. A., Chaffee, F. C., Foltz, C. B., Weymann, R. J., 2001, ApJ, 550, 142
\bibitem[Hamann et al.(2011)]{2011MNRAS.410.1957H} Hamann, F., Kanekar, N., 
Prochaska, J.~X., et al.\ 2011, \mnras, 410, 1957 

\bibitem[Harris et al.(2016)]{2016AJ.151..155} Harris, D. W., et al.\ 2016, \aj, 151, 155
\bibitem[He et al.(2014)]{2014MNRAS.443..2532} He, Z. C., Bian, W. H., Jiang, X. L., \& Wang, Y. F.\ 2014, \mnras, 443, 2532
\bibitem[He et al.(2015)]{2015MNRAS.454..3962} He, Z. C., Bian, W. H., Ge, X., \& Jiang, X. L.\ 2015, \mnras, 454, 3962
\bibitem[Hopkins et al.(2006)]{2006ApJS.163..1} Hopkins, P. F., Hernquist, L., Cox, T. J., Di Matteo, T., Robertson, B., Springel, V., 2006, ApJS, 163, 1
\bibitem[Hopkins et al.(2010)]{2010MNRAS.401..7} Hopkins, P. F., Elvis, M.\ 2010, \mnras, 401, 7
\bibitem[Kinney et al.(1990)]{1990ApJ...357...338K} Kinney, A., Rivolo, A.R., Koratkar, A.P.\ 1990, \apj, 357,338 
\bibitem[Konigl et al.(1994)]{1994ApJ...434..446} Konigl, A., \& Kartje, J. F.\ 1994, \apj, 434, 446
\bibitem[Leighly et al.(2015)]{2015ApJ...809...13}Leighly, K. M., Cooper, E., Grupe, D., Terndrup, D M., Komossa, S.,\ 2015, \apj, 809, 13
\bibitem[Liu et al.(2013)]{2013MNRAS...430...2327}Liu, G., Zakamska, N. L., Greene, J. E., Nesvadba, N. P. H., \& Liu,
X. 2013a, MNRAS, 430, 2327
\bibitem[Liu et al.(2013)]{2013MNRAS...436...2576}Liu, G., Zakamska, N. L., Greene, J. E., Nesvadba, N. P. H., \& Liu,
X. 2013b, MNRAS, 436, 2576
\bibitem[Liu et al.(2013)]{2014MNRAS...442...1303}Liu, G., Zakamska, N. L., \& Greene, J. E. 2014, MNRAS, 442, 1303
\bibitem[Liu et al.(2013)]{2015ApJS...221...9}Liu, G., Arav, N., \& Rupke, D. S. N. 2015, ApJS, 221, 9
\bibitem[Lundgren et al.(2007)]{2007ApJ...656...73L} Lundgren, B.~F., 
Wilhite, B.~C., Brunner, R.~J., et al.\ 2007, \apj, 656, 73
\bibitem[Moe et al.(2009)]{2009ApJ...706...525}Moe, M. et al.\ 2009, \apj, 706, 525
\bibitem[Moll et al.(2007)]{2007ApJ...463...513}Moll, R. et al.\ 2007, A\&A, 463, 513
\bibitem[P\^{a}ris et al.(2016)]{arXiv:1608.06483}P\^{a}ris, I. et al., 2016, arXiv:1608.06483
\bibitem[Riffel \& Storchi-Bergmann (2011)]{2011MNRAS...417...2752}Riffel, R. A., \& Storchi-Bergmann, T. 2011, MNRAS, 417, 2752
\bibitem[Rupke \& Veilleux (2013)]{2013ApJ...768.75}Rupke, D. S. N., Veilleux, S., 2013, ApJ, 768, 75
\bibitem[Scannapieco et al.(2004)]{2004ApJ...608.62} Scannapieco, E., \& Oh, S. P.\ 2004, \apj, 608, 62
\bibitem[Shen et al.(2011)]{2011ApJS..194...45S} Shen, Y., Richards, G.~T., Strauss, M.~A., et al.\ 2011, \apjs, 194, 45 
\bibitem[Sturm et al.(2011)]{2011EAS..52...55S} Sturm, E. et al.\ 2011, EAS, 52, 55S 
\bibitem[Tombesi et al.(2010)]{2010A&A...521A..57T} Tombesi, F., Cappi, M., Reeves, J. N., Palumbo, G. G. C., Yaqoob, T., Braito, V., Dadina, M. 2010, A\&A, 521A, 57T
\bibitem[Tombesi et al.(2011)]{2011ApJ...742..44T} Tombesi, F., Cappi, M., Reeves, J. N., Palumbo, G. G. C., Braito, V., Dadina, M. 2011, \apj, 742, 44T
\bibitem[Trevese et al.(2013)]{2013A&A...557A..91T} Trevese, D., Saturni, F.~G., Vagnetti, F., Perna, M., Paris, D., Turriziani, S.\ 2013, \aap, 557, A91
\bibitem[Wang et al.(2015)]{ 2015ApJ...814...150}Wang, T. G., Yang, C. W., Wang, H. Y., \& Ferland, G. \ 2015, \apj, 814, 150
\bibitem[Waters et al.(2016)]{2016arXiv161100407W} Waters, T., Proga, D., Dannen, R., \& Kallman, T.\ 2016, arXiv:1611.00407
\bibitem[Welling et al.(2014)]{2014MNRAS.440.2474W} Welling, C.~A., Miller, 
B.~P., Brandt, W.~N., Capellupo, D.~M.,\ 2014, \mnras, 440, 2474 
\bibitem[Weymann et al.(1991)]{1991ApJ...373...23W} Weymann, R.~J., Morris, 
S.~L., Foltz, C.~B., \& Hewett, P.~C.\ 1991, \apj, 373, 23 
\bibitem[Wildy et al.(2014)]{2014MNRAS.437.1976W} Wildy, C., Goad, M.~R., 
\& Allen, J.~T.\ 2014, \mnras, 437, 1976
\end{thebibliography}
\end{document}